\documentclass[twoside]{article}
\usepackage{qic,epsfig,amsmath,hyperref}
\usepackage{cite}
\usepackage{graphicx}
\usepackage{float}
\usepackage{multirow}

\renewcommand{\thefootnote}{\fnsymbol{footnote}}  

\newcommand{\fref}[1]{Fig.~\ref{#1}}

\newcommand{\tref}[1]{Table~\ref{#1}}
\newcommand{\sref}[1]{Section~\ref{#1}}

\AtBeginDocument{\hypersetup{pdfborder={0 0 0}}}

\newcounter{daggerfootnote}
\newcommand*{\daggerfootnote}[1]{%
    \setcounter{daggerfootnote}{\value{footnote}}%
    \renewcommand*{\thefootnote}{\fnsymbol{footnote}}%
    \footnote[2]{#1}%
    \setcounter{footnote}{\value{daggerfootnote}}%
    \renewcommand*{\thefootnote}{\arabic{footnote}}%
    }

\begin{document}
\setlength{\textheight}{8.0truein}    

\runninghead{Networked Quantum Services}
            {Laszlo Gyongyosi and Sandor Imre}

\normalsize\textlineskip
\thispagestyle{empty}
\setcounter{page}{1}

\copyrightheading{25}{2}{2025}{000--000}

\vspace*{0.88truein}

\alphfootnote

\fpage{1}

\centerline{\bf
NETWORKED QUANTUM SERVICES\daggerfootnote{Invited paper}}
\vspace*{0.035truein}
\vspace*{0.37truein}
\centerline{\footnotesize
LASZLO GYONGYOSI\footnote{Corresponding author}}
\vspace*{0.015truein}
\centerline{\footnotesize\it Department of Networked Systems and Services, Budapest University of Technology and Economics}
\baselineskip=10pt
\centerline{\footnotesize\it Budapest, H-1111, Hungary}
\vspace*{10pt}
\centerline{\footnotesize 
SANDOR IMRE}
\vspace*{0.015truein}
\centerline{\footnotesize\it Department of Networked Systems and Services, Budapest University of Technology and Economics}
\baselineskip=10pt
\centerline{\footnotesize\it Budapest, H-1111, Hungary}
\vspace*{0.225truein}
\publisher{(received date)}{(revised date)}

\vspace*{0.21truein}

\abstracts{
The intense growth of quantum computation and communication allows the development of advanced solutions and services. Networked quantum services are provided for the users via quantum computers and quantum networking. Here, we review the fundamental concepts and recent achievements of networked quantum services. We present a comprehensive study of the state of the art, the different technologies, platforms and applications. We analyze the implementation basis and identify key challenges.
}{}{}

\vspace*{10pt}

\vspace*{3pt}

\vspace*{1pt}\textlineskip    
\renewcommand{\thefootnote}{\arabic{footnote}}
\setcounter{footnote}{0}
\section{Introduction} 
One of the main drivers of today's scientific and technological development is the constantly growing computing power. However, we often run into the limitations of current computing technology. Keeping Moore's law alive is not possible by relying on classical physics. Although quantum computers open up new horizons, their full development will take a long time. Therefore, it is of utmost importance to combine the capabilities of machines with a limited number of qubits.

Quantum networked devices are systems that leverage quantum mechanics to establish communication networks, which can be used for applications involving security and privacy, remote sensing, or distributed computational capabilities, including the quantum cloud, quantum artificial intelligence, and solving computationally hard problems by means of limited-qubit computers.

Quantum sensing vastly improves the accuracy of how we measure and interact with the world around us by sensing changes in electric and magnetic fields. Collecting data at the atomic level means extracting information from individual atoms instead of from the huge collections of atoms, as usual in classical systems. This allows quantum sensors to make our devices exponentially more accurate. Devices that use quantum sensing are also not subject to the same physical constraints as conventional sensors, allowing for exceptional reliability with less vulnerability to the signal jamming and other electromagnetic interference that is increasingly common with today's light- and sound-based data sensors. Quantum sensors can be applied to build devices -- among others -- for faster, more accurate, more reliable geolocation, more detailed and accurate medical diagnostic images with fewer potential side effects for patients, reliable detection, imaging, and mapping of underground environments and deeper, more active sensing of gravitational changes and tectonic shifts that can forewarn or trigger avalanches, earthquakes, volcanic eruptions.

Distributed quantum computing represents a new computing paradigm in which quantum processing tasks are spread across multiple quantum devices or nodes, thus allowing these devices to work collaboratively on complex problems. In distributed quantum computing, qubits are managed across different locations or devices, which enables parallel processing of quantum algorithms. This approach can harness the combined power of multiple quantum systems. As the workload is distributed, systems can share computational resources, which is especially useful when individual quantum devices have limited qubit counts or processing capabilities. Entanglement plays a crucial role both in well-known efficient communication protocols \cite{Benett1,Benett2,Imre1} and algorithms \cite{Shor, Jozsa, Imre2}. Distributed quantum computing relies on entanglement between qubits located in different devices; this allows for quantum information to be shared and manipulated across the network, thereby enhancing computational power and efficiency.

The impact of quantum networks on real-world scenarios relies on the utilization of advanced quantum phenomena, such as quantum superposition and quantum entanglement, which prove useful in achieving performance better than classical networks. Quantum networks use quantum channels \cite{ieeecomm} to transmit the superposed or entangled quantum states (i.e., qubits in practical scenarios) from a sender user to a receiver user. The quantum channels can preserve the state of the qubits over long distances, allowing the development of efficient quantum networking services.

In the physical layer, the quantum channels can be implemented by different physical systems, such as optical fibers of current telecommunication systems, wireless optical channels, atoms, and superconducting technology. The quantum channels also allow high-speed communications similar to classical networking, but with significantly extended possibilities in data communications (quantum teleportation, superdense coding, distributed quantum computing, entanglement-based communication protocols) and advanced security via QKD (quantum key distribution) and quantum cryptographic protocols compared to classical networking. These fundamentally new possibilities make it possible to overcome several issues and drawbacks of classical telecommunications. In real-world scenarios, quantum channels and networking protocols are integrated into the existing classical telecommunication infrastructure. Classical channels are used to provide side information for the quantum operations of quantum networks. These data include controlling information, synchronization, signaling and timing for performing quantum operations. A distributed quantum protocol aims to perform some communication steps between the network users or complete a computational task in a distributed manner. The distant quantum computers and devices are connected via a set of intermediate quantum repeaters. The quantum repeaters establish a set of entangled connections with different entanglement levels between the users. The network structure integrates optical fiber and wireless optical channels between the ground-to-ground quantum repeaters and free-space optical channels for the ground-to-satellite communications.

However, in order to build a network from our currently available quantum devices, we still have to solve many technological problems. In order for these tools to work together, additional challenges must be faced.

\subsection{Novel Contributions} 
The novel contributions of our review article are as follows. 
\begin{itemlist}
\item We provide a comprehensive review of networked quantum services, with a detailed overview of the different technologies, platforms, and applications.
\item We study the core concepts of networked quantum services. We review recent advances in networked quantum services including networked quantum computers, networked quantum machine learning, networked quantum software, quantum programming languages, quantum cloud and quantum application programming interfaces. 
\item We present a comprehensive review of the state of the art of networked quantum services in a compact form and using well structured, easy-to-access tables.
\item We study the implementation basis of networked quantum services, and identify key challenges and open problems for future research. 
\end{itemlist}

\subsection{Structure of the Survey}
The structure of this survey is as follows. \sref{sec2} summarizes the related surveys. In \sref{sec3}, we comprehensively review the fundamental methods, and recent research on networked quantum services. We study the networked quantum devices, the architecture of networked quantum computers, and networked quantum machine learning. \sref{sec4} focuses on the quantum software of networked quantum services, the quantum programming languages, quantum application programming interfaces (APIs) and on the standardization approaches. \sref{sec5} reviews the implementation basis. Finally, \sref{sec6} concludes the survey, and outlines open problems and future research directions.

\section{Related Work} 
\label{sec2}
Recent surveys related to quantum computing, quantum communications and quantum software tools can be found in the literature.
Memon et al. \cite{qrep} reviewed the potential and current challenges of quantum computing along with important questions of quantum hardware development. Yang et al. \cite{survyang} summarized the current issues in quantum computing and communications focusing on quantum machine learning, quantum communications and quantum internet, quantum neural networks, and quantum cryptography. Sahu et al. \cite{nuts} reviewed the different quantum development tools and kits for quantum hardware and software. The survey of Singh et al. \cite{qprog3} also studied the available tools and technologies for quantum computing. Gyongyosi et al. \cite{surv1} reviewed the results of quantum computing technology. The survey of Gill et al. \cite{qcadd12} studies the taxonomy, and provides a systematic review on quantum computing.

Chae et al. \cite{nano} summarized the recent developments of qubits for quantum computing, while Bochkarev et al. \cite{discrete} studied the application of quantum computing for discrete optimization. The survey of Li et al. \cite{survli} focused on quantum optimization and quantum machine learning. Kusyk et al. \cite{survkus} discussed the methods of quantum circuit compilation for quantum computers of the NISQ (noisy intermediate scale quantum) era, along with the utilization of artificial intelligence methods in compilation. Upama et al. \cite{survupa} overviewed the different quantum programming languages, quantum simulators and quantum compilers. Massoli et al. \cite{survmas} focused on quantum computing and quantum neural networks. The work of Ramezani et al. \cite{survram} reviewed the different quantum machine learning algorithms. The survey of Peral-Garcia et al. \cite{mirev} presented a comprehensive overview and literature review on quantum machine learning. Tensor networks allow an efficient solution to simulate general quantum systems on classical computers and to study high-dimensional numerical problems. Garcia et al. \cite{tensorn} studied the computational applications of tensor network simulations. The survey of Ayral et al. \cite{manybody} studied the connections between quantum many-body systems and quantum processors. 

Moguel et al. \cite{survmog} reviewed the current state of quantum software engineering and service-oriented quantum computing. Serrano et al.  \cite{survser} studied quantum software components and platforms along with the quality assessment requirements for quantum software. The survey of Khan et al. \cite{survkhan} focused on the different architectures, frameworks, and tools for quantum software engineering. Jimnez-Navajas et al. \cite{qsdk2} reviewed the actual status of quantum software development, the current trends and requirements. The survey of Garhwal et al. \cite{qprog2} focused primarily on quantum programming languages. Dwivedi et al. \cite{lifec} reviewed quantum software engineering and the quantum software development lifecycle.

Cuomo et al. \cite{survcuo} and Caleffi et al. \cite{survcal} overviewed the current status of distributed quantum computing and quantum networking. Barrala et al. \cite{dcrev} presented a comprehensive overview of distributed quantum computing including the different quantum communication protocols and quantum networking aspects. The survey of Boschero et al. \cite{dcomp2} studied the different applications and challenges of distributed quantum computing. Jones et al. \cite{dcomp1} reviewed the utilization of distributed quantum computing for chemical applications. The survey of Pira et al. \cite{pira} reviewed the current state of the art in distributed quantum neural networks. Nguyen et al. \cite{cloudrev1} studied quantum cloud computing with the concepts, architecture models, software and resource management. In \cite{cloudrev2}, Moguel et al. reviewed the development and deployment of quantum services. The authors propose a comparison between different quantum computing service providers using a case study via empirical tests. 

In \cite{telecoms}, Phillipson studied the applications of quantum computing in telecommunications. The survey of Baseri et al. \cite{cyber} focused on the security aspects of quantum networking, while Dutta et al. \cite{commtrends} reviewed the recent trends and challenges of quantum communications. Li et al. \cite{nlayers} overviewed the fundamentals of entanglement-assisted quantum networks, while Mehic et al. \cite{meh} and Popa et al.\cite{pop} reviewed the applications of quantum cryptography in telecommunications. The survey of Abane et al. \cite{routings} proposed an comprehensive overview on the different entanglement routing methods in quantum networks. Gyongyosi et al. \cite{cacm} overviewed the fundamental attributes of the quantum internet, while in \cite{ieeecomm} Gyongyosi et al. presented a comprehensive overview on quantum channels. Wehner et al. \cite{survweh} studied the current stage of the quantum internet along with the requirements of a reliable large-scale global quantum communication network. Li et al. \cite{layered} studied the protocols of the quantum internet.

Currently, there are no surveys that specifically address networked quantum services. The studies discussed above also did not focus exhaustively on the related technologies of networked quantum services.

\subsection{Comparative Analysis}
For a comparative analysis, the primary scope of the related works are summarized in \tref{compar}.

\begin{table}[h!]
\footnotesize
\centering
\tcaption{Comparison of the primary scopes of the surveys.}
\label{compar}
\begin{tabular}{p{1.5in}p{3.6in}} \hline 
\centerline{Survey} & Primary scope \\ \hline 
Abane et al. \cite{routings} & Quantum internet.\\
Ayral et al. \cite{manybody}& Quantum computing.\\
Barrala et al. \cite{dcrev} & Distributed quantum computing.\\
Baseri et al. \cite{cyber} & Quantum communication.\\
Bochkarev et al. \cite{discrete} & Quantum computation.\\
Boschero et al. \cite{dcomp2} &	Distributed quantum computing.\\
Caleffi et al. \cite{survcal} &	Distributed quantum computing.\\
Chae et al. \cite{nano}	& Quantum computation.\\
Cuomo et al. \cite{survcuo} & Distributed quantum computing.\\
Dutta et al. \cite{commtrends} & Quantum communication.\\
Dwivedi et al. \cite{lifec} & Quantum development tools.\\
Garcia et al. \cite{tensorn} & Quantum computing, quantum machine learning.\\
Garhwal et al. \cite{qprog2} & Quantum programming languages.\\
Gill et al. \cite{qcadd12} & Quantum computing.\\
Gyongyosi et al. \cite{surv1} & Quantum computing.\\ 
Gyongyosi et al.  \cite{ieeecomm} & Quantum communication.\\
Gyongyosi et al. \cite{cacm} & Quantum internet.\\
Jimnez-Navajas et al. \cite{qsdk2} & Quantum programming languages, quantum development tools.\\
Jones et al. \cite{dcomp1} & Distributed quantum computing.\\
Khan et al. \cite{survkhan} & Quantum programming languages, quantum development tools.\\
Kusyk et al. \cite{survkus} & Quantum computation, quantum machine learning.\\
Li et al. \cite{layered} & Quantum internet.\\
Li et al. \cite{nlayers} & Quantum communication.\\
Li et al. \cite{survli} & Quantum computation, quantum machine learning.\\
Massoli et al. \cite{survmas} & Quantum computing, quantum machine learning.\\
Mehic et al. \cite{meh} & Quantum cryptography, quantum communication.\\
Memon et al. \cite{qrep} & Quantum computation.\\
Moguel et al.  \cite{cloudrev2} & Development and deployment of quantum services.\\
Moguel et al. \cite{survmog} & Quantum programming languages, quantum development tools.\\
Nguyen et al. \cite{cloudrev1} & Quantum cloud computing.\\
Peral-Garcia et al. \cite{mirev} & Quantum machine learning.\\
Phillipson \cite{telecoms} & Quantum computing, quantum communication.\\
Pira et al. \cite{pira} & Distributed quantum neural networks.\\
Popa et al.\cite{pop} & Quantum cryptography, quantum communication.\\
Ramezani et al. \cite{survram} & Quantum machine learning.\\
Sahu et al. \cite{nuts} & Quantum computation, quantum development tools.\\
Serrano et al. \cite{survser} & Quantum programming languages, quantum development tools.\\
Singh et al. \cite{qprog3} & Quantum computation, quantum programming languages.\\
Upama et al. \cite{survupa} & Quantum programming languages, quantum simulators.\\
Wehner et al. \cite{survweh} & Quantum internet.\\
Yang et al. \cite{survyang} & Quantum computation, quantum communication, quantum machine learning. \\\hline
\end{tabular}
\end{table}

\section{Architecture for Networked Quantum Services}
\label{sec3}
Networked quantum services represent a novel approach to significantly increase computational power compared to the computational capabilities of classical distributed services. The main idea behind networked services is the efficient utilization of quantum resources, such as quantum superposition and entanglement in distributed computations. The services could also use efficient quantum communications between the users and classical communications for the distribution of side information between the quantum nodes.

In practical scenarios, the quantum advantage of networked quantum services over classical distributed computations can be improved through the unconditionally secure quantum communications between the parties. Quantum entanglement applications in various protocols significantly extends the possibilities of classical networking. The fundamentals of quantum networking and secure quantum protocols, like QKD, provide the background for the advanced security of networked quantum services. The high-speed and high-fidelity quantum channels provide the reliable high-speed quantum data transmission for the distributed quantum computations. The scalability of networked services allows the use of quantum devices with a few qubits and quantum circuits for implementing efficient quantum computations in a distributed manner. 

The quantum internet \cite{cacm} provides a large-scale quantum network to implement networked quantum services in experiments. The quantum internet utilizes quantum channels (mainly optical fibers of free-space optical channels), quantum repeaters, quantum devices with quantum memory, and quantum computers to provide scalable high-speed networked services with unconditional security. The building blocks of the quantum internet extend the advanced properties of quantum communications to realize distributed services using quantum computer clusters (clusters of quantum devices with few quantum resources) and long-distance quantum communications. 

\tref{super} demonstrates the superior of networked quantum services to classical distributed computing with some practical applications of distributed quantum computing.  

\begin{table}[h!]
\footnotesize
\centering
\tcaption{Superiority of networked quantum services to classical distributed computing.}
\label{super}
\begin{tabular}{p{1.3in}p{2in}p{2in}} \hline 
\centerline{Reference} & \centerline{Result} & \centerline{Superiority}\\ \hline 
Ang et al. \cite{multinode} & Multinode quantum computing architecture. & Fast distributed computations. \\
Barz et al \cite{blinddem1}, Ruiting et al. \cite{blinddem2}, Mantri et al. \cite{blinddem3} & Demonstration of blind quantum computing. & Enhanced privacy. \\
Cirac et al. \cite{cirac2000} & Distributed phase estimation problem over noisy channels. & Fast distributed estimation of the phase of eigenvalues of a unitary operators.\\
Collins et al. \cite{collins2001} & Nonlocal content of quantum operations. & Local implementation of non-local quantum gates in a distributed quantum computer.\\
Eisert et al. \cite{eisert2000} & Resource-optimized protocols for non-local quantum gates. & Local implementation of non-local quantum gates in a distributed quantum computer.\\
Gidney et al. \cite{factor2} & Factoring RSA integers via distributed Shor algorithm. & Exponential speedup in distributed integer factorization and in discrete logarithm problems.\\
Gyongyosi et al. \cite{multiple} & Distributed multiple access QKD. & Enhanced security.\\ 
Gyongyosi et al. \cite{resbal} & Distributed resource allocation. & Improved resource prioritization and balancing.\\
Gyongyosi et al. \cite{scal} & Distributed problem-solving. & Fast distributed computations for optimization problems.\\
Li et al. \cite{ddj} & Distributed Deutsch-Jozsa algorithm. & Exponential speedup over distributed deterministic classical computers. \\
Neumann et al. \cite{dphase2} & Distributed quantum phase estimation. & Fast distributed estimation of the phase of eigenvalues of a unitary operators.\\
Nguyen et al. \cite{cloudrev1} & Quantum cloud computing. & Improved security, faster distributed computations.\\
Shi et al. \cite{dphase3} & Quantum message passing interface. & Fast distributed computations for statistical and optimization problems, constraint-satisfaction and graph isomorphism problems.\\
Tan et al. \cite{dsimon} & Distributed Simon's quantum algorithm. & Exponential speedup over distributed probabilistic classical computers. \\
Van Meter et al. \cite{vanmeter} & Arithmetic on a distributed quantum computer. & Exponential speedup in distributed integer factorization and in discrete logarithm problems.\\
Xiao et al. \cite{factor3} & Distributed quantum-classical factoring algorithm. & Exponential speedup in distributed integer factorization and in discrete logarithm problems.\\
Zhang et al. \cite{distrqsens}, Degen et al. \cite{sens1} & Distributed quantum sensing. & Improved accuracy in distributed environment sensoring and data acquisition.\\
Zhou et al. \cite{dbv} & Distributed Bernstein-Vazirani algorithm. & Efficient distributed solution of black-box problems.\\
Zhou et al. \cite{dgrov} & Distributed quantum searching algorithm. & Quadratic speedup in searching of elements in unstructured databases.\\
\hline
\end{tabular}
\end{table}

The core concept of networked quantum services is to use several smaller quantum processing units (QPUs) in parallel for the processing of input quantum data, then the output of the QPUs are combined via post-processing. In general, each QPU can contain a given number of smaller quantum circuits with limited quantum resources. The use of multiple QPUs allows a scalable architectural, since the connection of smaller distant QPUs in a distributed network allows greater processing capabilities. Networked quantum services can be classified according to their available quantum resources and communication channels. In all the different concepts, classical communication is always available between the distant QPUs for sharing side-information and for post-processing. However, quantum communication is not always available between the QPUs. As we will detail later, these circumstances lead to different conceptions and models of networked quantum services. 

\begin{figure} [!h]
\centerline{\includegraphics[angle = 0,width=1\linewidth]{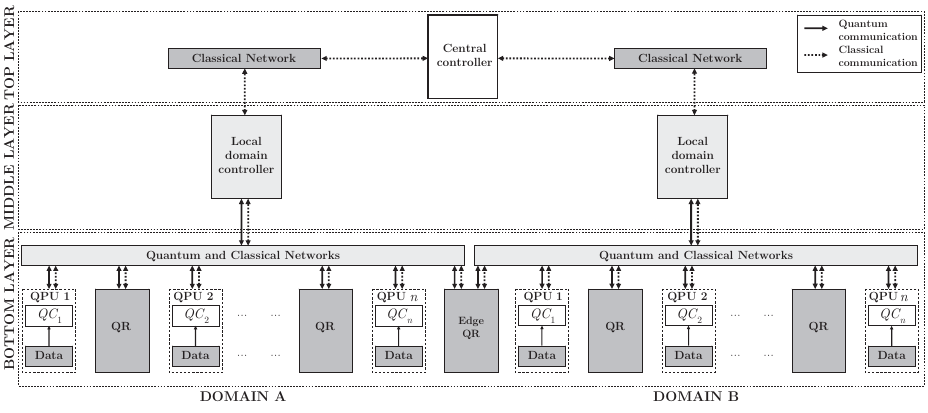}} 
\fcaption{A hierarchical quantum internet architecture for networked quantum services in compliance with the RFC 9340 standard \cite{archprinc,hierarch}. The $n$ QPUs, QPU $i$, $i=1,\ldots ,n$, process in parallel using their local input quantum data (Data) and local quantum circuits ${QC}_{i}$. The outputs of the QPUs are combined via post-processing. A given domain integrates $n$ QPUs, $n-1$ intermediate quantum repeaters (QR), and a local domain controller unit. The edge quantum repeater (Edge QR) logically separates the domains (Domains A and B). Classical and quantum communications are depicted by dashed and solid-lined arrows.}
\label{qint}
\end{figure}

The concept of the quantum internet offers the most favorable architectural conditions for networked quantum services. In a quantum internet setting, the distant QPUs have access to quantum communication channels and quantum resources for entanglement generation and distribution. The entanglement distribution procedure uses quantum repeaters between the distant QPUs and different protocols such as entanglement swapping. Quantum entanglement between the distant QPUs allow implementing different quantum communication protocols, such as quantum teleportation, quantum key distribution, superdense coding, etc. A hierarchical quantum internet architecture \cite{hierarch} for networked quantum services in compliance with the RFC 9340 \cite{archprinc} standard is depicted in \fref{qint}. The bottom layer integrates the QPUs and the quantum repeaters. The quantum devices communicate with a local domain controller via a quantum network and a classical network. This layer is also responsible for performing quantum operations, and it is referred to as the infrastructure plane. The middle and top layers are referred to as the control plane. The middle layer contains local domain controllers, which handle the entanglement generation and distribution between the QPUs. The top layer consists of a central controller, which uses a classical network for communication with the local domain controllers. The central unit controls the error correction\footnote{On a detailed description of quantum error correction, fault tolerance, and quantum memory limitations, we suggest references \cite{error1,error2,error3,error4}.} and the collection of global network information, and is responsible for a single quantum local area network (Q-LAN). Throughout our manuscript the classical and quantum communications are depicted by dashed and solid-lined arrows in the figures.

Since the general structure of a quantum internet is currently under development, the implementation of networked quantum services cannot assume that quantum entanglement between distant QPUs is available. Furthermore, the quantum circuits of the QPUs in the NISQ era are small circuits with a reduced number of qubits, called \textit{shallow} circuits \cite{shallow2}, in which the circuit depth increases no faster than logarithmically in the number of qubits. Shallow quantum circuits have a crucial significance in the NISQ era due to the limitations of current quantum hardware, such as the lack of error correction, limited resources, and high levels of disturbance and noise.

\subsection{Networked Quantum Devices}
The various technologies \cite{impl1,impl2,nuts} that are available for the implementation of quantum computations (such as photonic \cite{photonicqc}, superconducting \cite{supercond,qc2}, silicon-spins \cite{silic}, trapped neutral atom \cite{neutral}, giant atoms \cite{teleportrouter}, trapped ion \cite{ion}, topological \cite{topol}, nitrogen-vacancy \cite{nitrogen}, nuclear magnetic resonance-based \cite{nmr,nmr2} implementations) require a generalized framework and guidelines for the development of hardware for quantum computation. The aim of the layer architecture is to serve as an abstract hardware-independent standard for scalable fault-tolerant quantum hardware \cite{qclayer}.

The layer architecture of a general (gate-model) quantum device is as follows:
\begin{itemlist}
\item \textit{Physical Layer:} a hardware layer for physical quantum states, physical operations, quantum hardware and physical environment.
\item \textit{Virtual Layer:} conversion of physical systems to virtual primitives: virtual quantum states, virtual quantum gates, virtual quantum registers and virtual measurement. 
\item \textit{Error Correction Layer:} error correction and fault-tolerance methods, conversion of virtual primitives to logical primitives: logical quantum states, logical gates, logical quantum registers and logical measurement. 
\item \textit{Logical Layer:} logical abstractions for a universal quantum computer using the fault-tolerant logical structures.
\item \textit{Application Layer:} a software layer for interface with the user for data input and execution of quantum algorithms.
\end{itemlist}

Similar to the traditional OSI (Open Systems Interconnection) layer architecture, a layer architecture can also be defined for distributed quantum computation and networking with the aim of a hardware-independent standardization framework \cite{dcrev,rodr,vanmbook,dcomp1,dcomp2}.

The general layer architecture of distributed quantum computation consists of the following layers:
\begin{itemlist}
\item \textit{Physical Layer:} a hardware layer for physical mechanisms to connect distant nodes (quantum entanglement generation and distribution, entanglement swapping, entanglement purification, quantum teleportation, specific operations \cite{vanmbook}).
\item \textit{Network Layer:} a hardware layer for definition of procedures for establishing communications between distant quantum nodes.
\item \textit{Development Layer:} a software layer for mechanisms that allow the applications to be distributed and executed on the physically distributed quantum network (qubit mapping methods, partitioning, compilation, machine code generation, optimization \cite{rodr}).
\item \textit{Application Layer:} a software layer for user interface for data input and execution of distributed quantum algorithms.
\end{itemlist}

\begin{figure} [!h]
\centerline{\includegraphics[angle = 0,width=0.75\linewidth]{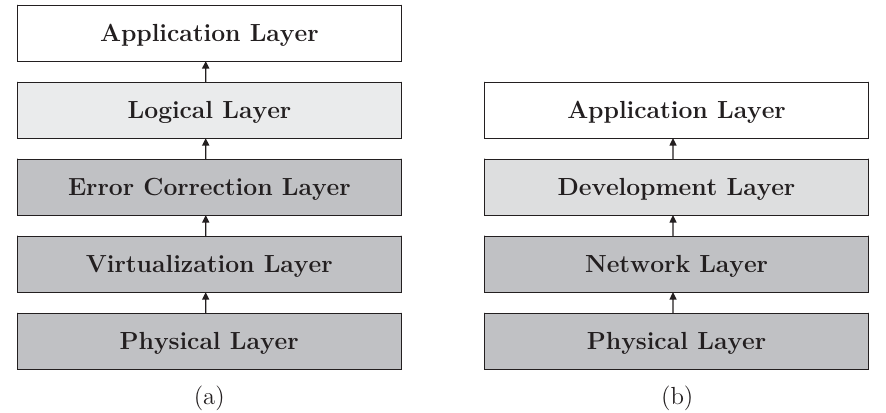}} 
\fcaption{\label{motion} Layer architectures. (a) Layer architecture of a quantum computer (general gate-model quantum device). (b) A general layer architecture for distributed quantum computation.}
\label{layers2}
\end{figure}

The layer structure of a quantum computer (gate-model quantum device) and the layered model for a distributed quantum computation are shown in \fref{layers2}.

While the layer structure can referred to as a general layer architecture, we note that other layer approaches are also available for distributed quantum computation \cite{nlayers,nlayers2,cacm,dcrev,archprinc}. The main attributes of the different approaches to layer architectures are summarized in \tref{layers}. 

\begin{table}[!h]
\footnotesize
\centering
\tcaption{Different layer architectures for distributed quantum computation.}
\label{layers}
\begin{tabular}{p{1.8in}p{0.4in}p{2.8in}} \hline 
\centerline{Approach} & \centerline{No. of layers} & \centerline{Layer architecture} \\ \hline 
Quantum networking via bipartite entanglement \cite{bipart1,bipart2} & \centerline{5} & Physical Layer, Link Layer, Network Layer, Transport Layer, Application Layer. \\
Quantum networking via multipartite entanglement \cite{pirklayer} & \centerline{4} & Physical Layer, Connectivity Layer, Link Layer, Network Layer. \\
Quantum recursive network architecture \cite{vanmlayer1,vanmlayer2} & \centerline{5} & Physical Layer, Link Layer, Remote State Composition Layer, Error Management Layer, Application Layer. \\
Multinode quantum computing architecture \cite{multinode}
& \centerline{6} & Physical Layer, Distillation Layer, Data Link Layer, Network Layer, Transport Layer, Application Layer. \\ \hline
\end{tabular}
\end{table}

\subsection{Networked Quantum Computers}
Quantum computing utilizes the fundamental concepts of quantum mechanics for computing, such as superposition, entanglement, and interference \cite{qc1,qc2,qc3,qc4,qc5,mas}. As the development of quantum computers evolve \cite{qc1,qc2,qc3,qcadd1,qcadd2,qcadd3,qcadd4,qcadd5,qcadd6,qcadd7,f1,f2,qc5,basis4,qcadd9,qcadd10,qcadd11,qcadd12,qcadd13,qcadd14,qcadd15,qcadd8}, the relevance of quantum computations has become more interpretable for problem-solving tasks \cite{qc2,circs,boson,boson2,boson3}. Quantum computers could definitely help solving difficult computational problems \cite{pnas,tech,tech2,yama,speedup,qc1,scottmi,surv1,nisq}.

As a first step, Shor's factorization algorithm \cite{fact} has been implemented via a small NMR (nuclear magnetic resonance) device in 2001 \cite{nmr}. In 2010, the quantum annealing computers has been announced by D-Wave\footnote{D-Wave. URL: \url{https://www.dwavesys.com}}, with limited application possibilities. The quantum computing research at Google has been started in 2013\footnote{Google Quantum AI. URL: \url{https://quantumai.google/}}. In 2016, they developed a simulated annealing algorithm \cite{anneal}, and also simulated the energy levels of the hydrogen molecule \cite{molec}. In 2019, Google has announced a 53-qubit gate-model superconducting quantum computer \cite{qc2}, which, in theory, can solve optimization problems in only 3 min that would take a classical supercomputer 10 000 years. A theoretical framework of the experiments of the Google's 53-qubit quantum computer has been defined in \cite{qc4}. Quantum computing research at IBM has been started in 2016\footnote{A new way of thinking: The IBM quantum experience. URL: \url{http://www.research.ibm.com/quantum}} ~via the IBM quantum experience program. IBM has made possible for users to utilize the IBM Quantum Cloud Services\footnote{IBM Cloud, IBM. URL: \url{https://www.ibm.com/cloud}} ~to run algorithms on the IBM quantum processor from the users' computers. Between 2016 and 2019, IBM has developed 5-qubit and 27-qubit quantum processors, while in 2020, they announced the 65-qubit IBM Quantum Hummingbird processor, a 127-qubit processor in 2021, and a 433-qubit processor in 2022. The 1121-qubit processor has been released in 2023\footnote{IBM Quantum Roadmap. URL: \url{https://research.ibm.com/blog/ibm-quantum-roadmap}}. In 2020, Amazon has launched the Amazon Braket Cloud Quantum Computing Service\footnote{{Amazon Braket. URL: \url{https://aws.amazon.com/braket/}}} ~to provide quantum-based web services.

In the NISQ era \cite{qc1,nisq}, gate-model quantum computers have particular relevance since these architectures can be implemented on near-term settings \cite{qc1,qc2,qc3,qc4,qc5,qcadd1,qcadd2,qcadd3,qcadd4,qcadd5,qcadd6,qcadd6b,qcadd6c,qcadd12,qcadd13,qcadd14,qcadd15,qcadd7,qcadd8,qcadd9,qcadd10,qcadd11,f1,f2,f3,f4}. In a gate-model quantum computer, the computational steps are realized via unitary gates. The gates are associated with a gate parameter value, while the computational problem fed into the quantum computer identifies an objective function \cite{qc5,basis4,f1,f2,f3,f4}. For objective function examples, see \cite{f3,f4,qc5,qcadd1,qcadd4,qcadd5}. The aim of the problem solving is to maximize the objective function value via several iteration steps. Each iteration step includes the application of unitary gates, as well as a measurement of the resulting quantum states. From the measurement results, an averaged value can be determined to estimate the actual objective function value \cite{qc5}. Many hybrid variational quantum algorithms also use gate-model circuits and classical objective functions \cite{var1,var2,pqc2,var4,ujv2_4}, similar to the quantum approximate optimization algorithm (QAOA) \cite{f1,qc5}.

The NISQ quantum devices are noisy, the decoherence and the imperfections of gates and measurements could easily destroy any possible quantum speedup on these devices. The lack of error-correction prevents \textit{scaling}. Fault-tolerance requires precise physical control over the qubits (two-qubit gates with 99.99\% fidelity) \cite{qc1,qrep}. Another crucial problem of quantum computing is scalability. As the number of qubits and the number of quantum gates increase, so do the errors and noise. \textit{Networked quantum computers} allow us to overcome the limitations of the current NISQ quantum hardware. In a networked setting, the quantum computations are realized by clusters of quantum nodes in a distributed manner. The aim of a distributed quantum computer architecture is to use multiple smaller quantum computers (equipped with a small number of qubits and quantum gates) instead of a single large quantum computer to perform the same computational task. The total numbers of qubits and gates of the networked architecture can exceed the numbers of qubits and gates of the large quantum computer, allowing efficient scalability with low error rates and noise levels. The interconnection between the quantum nodes requires control mechanisms in the hardware and software layers.

In a networked quantum computer architecture, the remote quantum processing units (QPUs) use classical and quantum channels to communicate with the classical network and quantum network. The quantum channel can be an optical link, free space optical channel, or other physical implementation (see also \sref{sec5}). In a large quantum computer, the levels of error and noise are high, and the coherence times are limited. In a distributed architecture, the QPUs are equipped with a significantly fewer qubits and quantum gates, which reduces the errors and noise, and overcomes the problem of the limited coherence times. It allows using simplified, low-complexity error-correction techniques in the nodes, and decreases the noise that arises from the imperfections of the quantum gates. Both attributes are crucial in an NISQ setting, where the quantum hardware and quantum resources are limited. 

We discuss networked quantum computers using the following categorization:
\begin{itemlist}
\item \textit{Multichip}: an input computational problem is divided into several smaller quantum circuits for parallel processing. The quantum circuits can work independently without the need for direct communication between them. The smaller quantum circuits fit in one QPU, no circuit decomposition is needed. The quantum communication between the quantum circuits is implementation-specific, and classical communication is always available. If only classical communication is available, it is also called \textit{embarrassing parallelism} \cite{dcrev} (\tref{multichip}).
\item \textit{Circuit distribution}: a large quantum circuit requires a decomposition into smaller quantum circuits. The smaller circuits are mapped to distant QPUs (nodes) using quantum communications, quantum resources and classical communications. Quantum and classical communication is available between the nodes (\tref{cdist}).
\item \textit{Circuit splitting}: a large quantum circuit is split into smaller quantum circuits between distant quantum nodes. Quantum communication is not available, classical communication is available (\tref{tab3}).
\end{itemlist}

The distributed architecture of networked quantum services allows the upscaling of the computational power of smaller quantum devices to a high computational capacity. In the multichip approach, several smaller quantum chips are used in parallel to realize a scalable quantum system. In the circuit distribution and splitting methods, a large quantum circuit of a large quantum computer is split into several smaller subcircuits and implemented in the distant quantum nodes. The splitting of the QC can be made in different ways: horizontally or vertically, in general. Note, the circuit splitting can refer to gate-cutting \cite{incoh2} or wire-cutting \cite{incoh1}, both can be shown to be equivalent \cite{eqiv}.
The main attributes of the multichip approach with the related works are summarized in \tref{multichip}. 

\begin{table}[!h]
\footnotesize
\centering
\tcaption{Multichip approaches.}
\label{multichip}
\begin{tabular}{p{1.4in}p{3.1in}} \hline 
\centerline{Attribute} & \centerline{Description and related works} \\ \hline 
\multirow{4}{*}{Multichip} 
& \multicolumn{1}{p{3.1in}}{Smaller quantum circuits are executed in parallel, outputs of the quantum circuits are combined via post-processing \cite{dcrev}.} \\
& \multicolumn{1}{p{3.1in}}{\textbf{Workload distribution:} the QPUs are scheduled for the computation of a subset of an input problem \cite{var1, wload1, wload2, wload3, pqc11}.} \\
& \multicolumn{1}{p{3.1in}}{\textbf{Offloading:} execution of programs with quantum tasks that are offloaded to a given QPU \cite{offload1,offload2,offload3,offload4,offload5}.} \\
& \multicolumn{1}{p{3.1in}}{\textbf{Mapping:} quantum circuits are physically mapped to the QPUs \cite{qmapping,qmapping2,qmapping3}.} \\
Quantum and classical communication & Quantum communication is implementation-specific, classical communication is available. \\
Applications & Phase estimation \cite{embphase}, amplitude estimation \cite{embamp}, quantum searching \cite{emapp1,emapp2}, multi-programming of quantum computers \cite{multip,multip2,multip3,multip4,interl}. \\
NISQ compatibility & Partial compatibility: multichip approaches with no entanglement between the circuits, due to the current quantum hardware and quantum resource limitations. \\ \hline 
\end{tabular}
\end{table}

The details of the circuit distribution method with the related works are given in \tref{cdist}. 

\begin{table}[!h]
\footnotesize
\centering
\tcaption{Circuit distribution approaches.}
\label{cdist}
\begin{tabular}{p{1.4in}p{3.1in}} \hline 
\centerline{Attribute} & \centerline{Description and related works} \\ \hline 
\multirow{4}{*}{Circuit distribution} 
& \multicolumn{1}{p{3.1in}}{\textbf{Partitioning:} the large quantum circuit is mapped onto a graph, and the optimal partitioning of the graph has to be find with minimal teleportation or distributed (non-local) gates between the nodes \cite{cd1,dcrev}.} \\
& \multicolumn{1}{p{3.1in}}{\textbf{Partitioning methods for quantum circuits:} deep reinforcement learning (DRL) \cite{drl}, Karlsruhe hypergraph partitioning \cite{karl1,karl2,cd1,karl3,karl4,karl5,karl6,karl7,karl8}, Kernighan-Lin partitioning \cite{kl,kl1,kl2,kl3,kl4}, Fiduccia-Mattheyses algorithm \cite{fid1,fid2}, tree-based directed acyclic graph (TDAG) \cite{tdag}, relaxed-overall extreme exchange (rOEE) \cite{oee1,oee2}, Hungarian qubit assigment (HQA) \cite{hunq}, quadratic unconstrained binary optimization (QUBO) \cite{qubo}.} \\
& \multicolumn{1}{p{3.1in}}{\textbf{Distribution of partitions:} via entanglement generation and quantum communications between the QPUs.} \\
& \multicolumn{1}{p{3.1in}}{\textbf{Partition mapping to QPUs:} the subcircuit is mapped to the physical structure in the QPU via a local map. Primary method: Fine Grained Partitioning (FGP) \cite{oee2,fgp}.} \\
Quantum and classical communication & Quantum communication is available, classical communication is available. \\
Applications & Distributed quantum gates \cite{vanmeter}, estimating the mean of numbers \cite{grovertelec}, distributed Simon's algorithm \cite{dsimon}, distributed quantum phase estimation \cite{cirac2000,dphase2,dphase3}, distributed Deutsch-Jozsa algorithm \cite{ddj}, distributed Bernstein-Vazirani algorithm \cite{dbv}, distributed quantum searching \cite{gupta,dgrov}, distributed quantum Fourier-transform \cite{dcnot}, distributed integer factoring \cite{factor1,factor2,factor3}.\\
NISQ compatibility & Partial compatibility, due to the current limitations of quantum hardware and quantum resources. \\ \hline 
\end{tabular}
\end{table}

Relevant attributes of the circuit splitting method with the related works are included in \tref{tab3}. 

\begin{table}[!h]
\footnotesize
\centering
\tcaption{Circuit splitting approaches.}
\label{tab3}
\begin{tabular}{p{1.4in}p{3.1in}} \hline 
\centerline{Attribute} & \centerline{Description and related works} \\ \hline 
\multirow{5}{*}{Circuit splitting} 
& \multicolumn{1}{p{3.1in}}{\textbf{Horizontal splitting:} a quantum circuit is split horizontally between distant nodes \cite{incoh1,incoh2,eqiv}.} \\
& \multicolumn{1}{p{3.1in}}{\textbf{Horizontal, incoherent splitting:} the quantum circuit is simulated by a sequence of local circuits followed by a quantum measurement of the qubits \cite{incoh1,incoh2, incoh3,incoh4, incoh5,incoh6, incoh7,incoh8}.} \\
& \multicolumn{1}{p{3.1in}}{\textbf{Horizontal, coherent splitting:} the quantum circuit is simulated by a sequence of local circuits with no quantum measurement on the qubits to preserve quantum information \cite{incoh2,coh1,coh2,coh3}.} \\ 
& \multicolumn{1}{p{3.1in}}{\textbf{Horizontal, combined incoherent and coherent splitting:} combination of incoherent and coherent splitting \cite{comb1}.} \\
& \multicolumn{1}{p{3.1in}}{\textbf{Vertical splitting:} a quantum circuit is split vertically between distant nodes.} \\
Quantum and classical communication & Quantum communication is not available, classical communication is available. Horizontal split requires classical communication between the nodes \cite{incoh8, coh2, comb1, coh1,horuj}, vertical split requires quantum tomography \cite{incoh5}. Utilization of quantum datasets \cite{coh1}. \\
Applications & Simulation of large quantum circuits \cite{incoh1}, optimal quantum circuit cuts to clustered Hamiltonian simulation \cite{harr25}, combinatorial optimization \cite{incoh7,combopt}, solution of chemical problems \cite{incoh4}, complex problem solving via small quantum circuits \cite{deepvqe}, high dimensional quantum machine learning \cite{highml}. \\
NISQ compatibility & Partial compatibility: horizontal, incoherent splitting due to the current limitations of quantum hardware. \\ \hline 
\end{tabular}
\end{table}

\begin{figure} [!h]
\centerline{\includegraphics[angle = 0,width=1\linewidth]{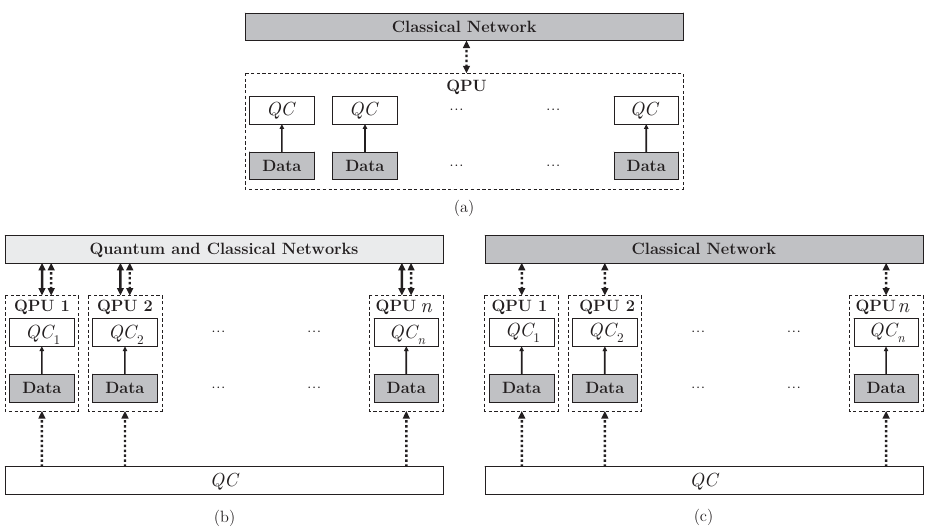}} 
\fcaption{\label{motion} Distributed quantum computer architectures. (a) Multichip: the $n$ smaller quantum circuits ($QC$) fit in a quantum processing unit (QPU), quantum communication is implementation-specific, classical communication is always available via Classical Network. (b) Circuit distribution: a large quantum circuit ($QC$) is distributed among the $n$ QPUs such that each node realizes a smaller subcircuit of it with relation $\sum\nolimits_{i=1}^{n}{{{w}_{i}}}{QC}_{i}=QC$, where ${QC}_{i}$ is the local circuit of the $i$-th QPU, while ${w}_{i}$ is a real weighting coefficient. Quantum communication is available (via Quantum Network), classical communication is available (via Classical Network). (c) Circuit splitting approach: a large quantum circuit ($QC$) is split among the $n$ QPUs, such that each node realizes a smaller subcircuit with relation $\sum\nolimits_{i=1}^{n}{{{w}_{i}}}{QC}_{i}=QC$, no quantum communication is available, classical communication is available.}
\label{dqarch}
\end{figure}

The multichip, circuit distribution and circuit splitting architectures are depicted in \fref{dqarch}.

\begin{figure} [!h]
\centerline{\includegraphics[angle = 0,width=0.5\linewidth]{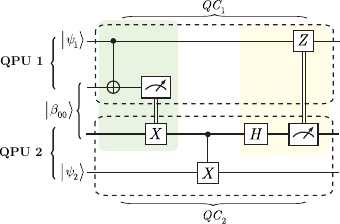}} 
\fcaption{\label{motion} A distributed CNOT gate realized by two QPUs between distant quantum states ${| \psi _{1} \rangle} $ and ${| \psi _{2} \rangle} $. Local system ${| \psi _{1} \rangle} $ is at QPU 1, while ${| \psi _{2} \rangle} $ is at QPU 2. QPU 1 has the quantum circuit $QC_{1} $ (dashed line frame) and QPU 2 has $QC_{2} $ (dashed line frame). The quantum nodes also share a Bell state ${| \beta _{00} \rangle} ={\textstyle\frac{1}{\sqrt{2} }} ({| 00 \rangle} +{| 11 \rangle} )$. The QPUs use their local quantum circuits and classical communication (doubled lines) to realize the CNOT gate between ${| \psi _{1} \rangle} $ and ${| \psi _{2} \rangle} $ in a distributed manner. The green-shaded box is the cat-entangler sequence \cite{dcnot}, the yellow-shaded box is the cat-disentangler sequence. Related implementations: \cite{dcnotimp1,dcnotimp2,dcnotimp3,dcnotimp4,dcnotimp5}.}
\label{dcnot}
\end{figure}

A distributed CNOT gate between two distant QPUs is shown in \fref{dcnot}.

Recent implementations of multichip and circuit splitting approaches are given in \tref{tabmultisplit}. 

\begin{table}[!h]
\footnotesize
\centering
\tcaption{Recent distributed quantum computing approaches and implementations.}
\label{tabmultisplit}
\begin{tabular}{p{0.9in}p{3.5in}} \hline
\centerline{Approach} & \centerline{Description and related works} \\ \hline
\multirow{7}{*}{Multichip} 
& \multicolumn{1}{p{3.5in}}{Architecture for multicore quantum computers with double full-stack communication \cite{rodr}.} \\
& \multicolumn{1}{p{3.5in}}{Quantum data networking for distributed quantum computing \cite{qiao}.} \\
& \multicolumn{1}{p{3.5in}}{Multi-qubit generation, development of multichip quantum computing platform \cite{cho}.} \\
& \multicolumn{1}{p{3.5in}}{Scalable multichip quantum architecture using hybrid wireless/quantum-coherent network \cite{alar}.} \\
& \multicolumn{1}{p{3.5in}}{Multichip multidimensional quantum network with entanglement retrievability \cite{zheng}.} \\
& \multicolumn{1}{p{3.5in}}{Modular superconducting-qubit architecture with a multichip tunable coupler \cite{field}.} \\
& \multicolumn{1}{p{3.5in}}{Variational quantum algorithms \cite{wload4,wload5,wload6,wload7,wload8,wload9}.} \\ \hline
\multirow{11}{*}{Circuit distribution}
& \multicolumn{1}{p{3.5in}}{Arithmetic on a distributed-memory quantum multicomputer \cite{vanmeter}.} \\
& \multicolumn{1}{p{3.5in}}{Scalable distributed gate-model quantum computers for distributed problem-solving \cite{scal}.} \\
& \multicolumn{1}{p{3.5in}}{Architectures for multinode superconducting quantum computers \cite{multinode}.} \\
& \multicolumn{1}{p{3.5in}}{Modular quantum compilation framework for distributed quantum computing \cite{ferr}.} \\
& \multicolumn{1}{p{3.5in}}{Factoring 2048 bit RSA integers using 20 million noisy qubits \cite{factor2}.} \\
& \multicolumn{1}{p{3.5in}}{Distributed quantum-classical hybrid factoring algorithm \cite{factor3}.} \\
& \multicolumn{1}{p{3.5in}}{Imperfect distributed quantum phase estimation \cite{dphase2}.} \\
& \multicolumn{1}{p{3.5in}}{Implementation for a quantum message passing interface \cite{dphase3}.} \\
& \multicolumn{1}{p{3.5in}}{Distributed quantum algorithm for Simon's problem \cite{dsimon}.} \\
& \multicolumn{1}{p{3.5in}}{Distributed quantum algorithms for Deutsch-Jozsa problem \cite{ddj}.} \\
& \multicolumn{1}{p{3.5in}}{Distributed Bernstein-Vazirani algorithm \cite{dbv}.} \\
& \multicolumn{1}{p{3.5in}}{Distributed Grover's algorithm \cite{dgrov}.} \\ \hline
\multirow{9}{*}{Circuit splitting}
& \multicolumn{1}{p{3.5in}}{Scalable quantum computing infrastructure using superconducting electronics \cite{circspl1}.} \\
& \multicolumn{1}{p{3.5in}}{Distributed quantum computation with circuit splitting \cite{incoh7}.} \\
& \multicolumn{1}{p{3.5in}}{Quantum circuit cutting with maximum-likelihood tomography \cite{incoh5}.} \\
& \multicolumn{1}{p{3.5in}}{A smart quantum circuit cutting method \cite{circspl3}.} \\
& \multicolumn{1}{p{3.5in}}{Hypergraphic partitioning of quantum circuits for distributed quantum computing \cite{fid2}.} \\
& \multicolumn{1}{p{3.5in}}{Fast quantum circuit cutting with randomized measurements \cite{incoh8}.} \\
& \multicolumn{1}{p{3.5in}}{Clifford-based circuit cutting for quantum simulation \cite{circspl5}.} \\
& \multicolumn{1}{p{3.5in}}{Scalable emulation of quantum algorithms on high-performance computers \cite{circspl6}.} \\
& \multicolumn{1}{p{3.5in}}{Dimensionality reduction via circuit splitting for quantum reinforcement learning \cite{circspl7}.} \\ \hline
\end{tabular}
\end{table}

\subsection{Networked Quantum Machine Learning}
In this section, we are focusing on distributed quantum neural networks (QNN). Quantum neural networks are also referred to as parameterized quantum circuits (PQC) or variational quantum circuits (VQC) \cite{qnnimp9,pqc1,pqc2,pqc3,pqc4,pqc5,nisq,pqc7, pqc8}.

The most promising approach for quantum machine learning in the NISQ setting is to use variational quantum algorithms (VQAs) \cite{ref31,qcadd1,scottmi}: this includes classical hardware to overcome the limitations of the current quantum hardware. VQAs are useful for machine learning problems \cite{ref31,qcadd1,scottmi} and combinatorial optimization \cite{var5,var6}. A VQA consists of a feature map, an architecture (ansatz), loss function and an optimization method \cite{pqc4, variat2}. These attributes make them a strong candidate for achieving a quantum advantage in the near term. The feature map encodes the classical input data into quantum states \cite{ham1,pqc4}. The algorithms use an optimizer on classical hardware to train a PQC, which is used to find the quantum state containing the solution to the problem \cite{pqc4,pqc}. However, the optimal choice of the optimizer, the structure of the PQC, and the hyperparameters are problem-specific and have a major impact on the performance of a VQA.

\begin{figure} [!h]
\centerline{\includegraphics[angle = 0,width=0.5\linewidth]{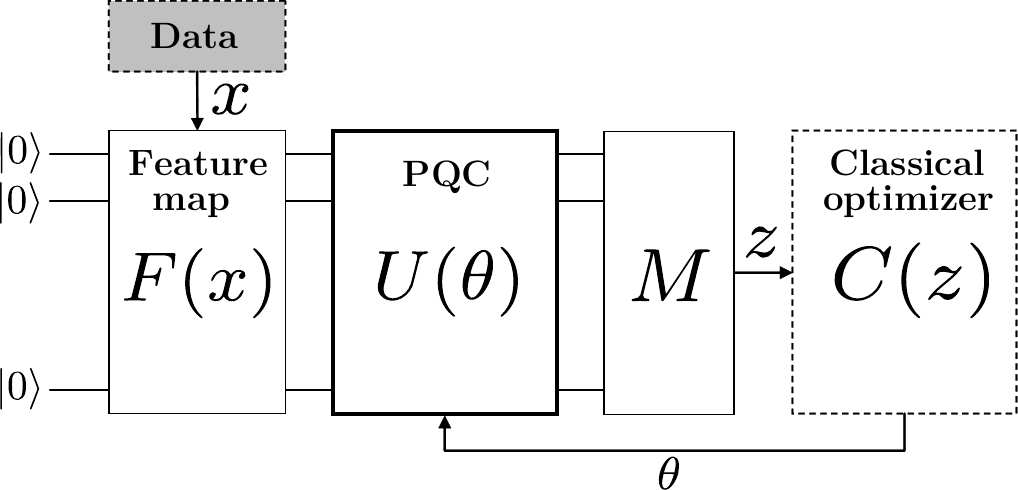}} 
\fcaption{\label{motion} Schematic model of a VQA. The classical data $x$ is mapped onto $n$ qubits via feature map $F(x)$, and fed into the parameterized quantum circuit $U(\theta )$ (PQC), where $\theta $ is the parameter of unitary $U$. The output of $U(\theta )$ is measured via measurement $M$ in the computational basis, which results in a classical bit string $z\in \{0,1\}^{n}$ of length $n$. This string is fed to a classical computer to evaluate a cost function $C(z)$. Depending on the value of $C(z)$, the parameter $\theta $ is updated via a classical optimizer and propagated back to $U(\theta )$. The circuit is re-run multiple times until a convergence or other stopping criterion is fulfilled.}
\label{vqa}
\end{figure}

A schematic model of a VQA is depicted in \fref{vqa}.

In general, quantum circuits are sequences of unitaries each of which depends on a few parameters. Due to the lack of error correction in the NISQ era, the number of gates that can be applied is limited by the fidelities of the quantum gates, and the tolerable error is low, which is a challenge for the implementation of gate-model quantum neural networks in near-term devices \cite{mirev,ref31,basis4,scottmi}. Depending on whether classical or quantum data is used, and a classical or quantum algorithm, the quantum machine learning architectures can be classified into different groups \cite{group,pira,pira2}. These groups with their main attributes and the related works are summarized in \tref{tab1}.

\begin{table}[!h]
\footnotesize
\centering
\tcaption{Classes of quantum machine learning based on the data type (classical or quantum) and the algorithm (classical or quantum).}
\label{tab1}
\begin{tabular}{p{0.5in}p{2in}p{2in}} \hline
\centerline{Data} & \multicolumn{2}{p{3.3in}}{\centerline{Algorithm}} \\ \hline 
& \textbf{Classical} & \textbf{Quantum} \\
\textbf{Classical} & Classical-Classical: classical data and classical machine learning algorithms inspired by quantum mechanics \cite{cc2}. & quantum-classical: classical data encoded into quantum states and processed by quantum processors for quantum speedups \cite{cq, qram1,qram2,qram3}, NISQ applications \cite{basis4, distr2}. \\
\textbf{Quantum} & Quantum-Classical: quantum data augmented by classical computations \cite{nqc1,nqc2,nqc3,nqc4,nqc5,nqc6,nqc7,nqc8}. & Quantum-Quantum: quantum data on quantum processors \cite{qq1,qq2}. \\ \hline 
\end{tabular}
\end{table}

Major challenges in the implementation of VQAs are trainability, mainly by the occurrence of the \textit{barren plateau} \cite{barren1,qnnbarr1,qnnbarr4,pqc4} and \textit{narrow gorge} \cite{gorge}, efficiently calculable cost functions and expectation values, \textit{multiple local minima} in the loss function \cite{local}, high precision measurements \cite{pqc4}, and ensuring high accuracy on near-term quantum processors with high levels of noise and errors. The hyperparameters of VQAs are the configuration parameters, such as those for the optimizer or the PQC \cite{pqc4,pqc,dipl1,dipl2}. It may be assumed that a further quantum advantage can be reached in terms of space efficiency, performance, and expressiveness of learning models in the NISQ era. Translational and exploratory approaches are studied in \cite{approach}. Different types of quantum neural networks have been defined in the literature, such as orthogonal, convolutional, feed-forward, self-supervised quantum neural networks, variational depth quantum circuits, quantum Boltzmann machines, along with the applications of linear regressions or quantum amplitude estimation. For a comprehensive review of quantum machine learning applications, see \cite{mirev}. 

In a distributed quantum neural network, the training is realized in a parallelized way between distant nodes. The parallelization can be made via \textit{data splitting} (\tref{tab2}) or \textit{circuit splitting} (\tref{tab3}). In data splitting approaches, the dataset is split across the quantum nodes such that the actual model is loaded into each node (term ``model'' refers to a PQC in this context). In circuit splitting, the model (a PQC) is split across the distant nodes such that each node has access to the full dataset. 

The main attributes of data splitting approaches (encoding, data storage and use of classical side information) in distributed quantum neural networks is summarized in \tref{tab2}.

\begin{table}[!h]
\footnotesize
\centering
\tcaption{Data splitting approaches in distributed quantum neural networks.}
\label{tab2}
\begin{tabular}{p{1.4in}p{3.1in}} \hline
\centerline{Attribute} & \centerline{Description and related works} \\ \hline 
\multirow{4}{*}{Data encoding} 
& \multicolumn{1}{p{3.1in}}{Encoding of a classical dataset into quantum states.}\\
& \multicolumn{1}{p{3.1in}}{\textbf{Basis encoding:} the data is encoded in a computational basis state. The dataset is encoded as superposition of the computational basis states \cite{basis0,basis1,rose}.} \\
& \multicolumn{1}{p{3.1in}}{\textbf{Amplitude encoding:} uses amplitudes of the quantum state for dataset encoding \cite{amp1,amp2}.} \\ 
& \multicolumn{1}{p{3.1in}}{\textbf{Angle encoding:} tensor product encoding at the single-qubit level without entanglement within feature vectors \cite{angle1,angle2,angle3,angle4}.} \\
& \multicolumn{1}{p{3.1in}}{\textbf{Hamiltonian encoding:} encoding is made at the Hamiltonian level mostly by two-qubit entangling gates \cite{ham1,var2,ham3,ham4}.} \\ 
Data storage & Quantum random access memory \cite{qram1,qram2,qram3}. \\ 
Quantum and classical communication & Quantum communication is available, quantum simulators are available, quantum AI software packages \cite{amp2}, classical communication is available \cite{weig1,weig2}. \\ 
Applications & Classification problems \cite{basis2, basis4, distr2}, quantum data compression \cite{basis3}, quantum Boltzmann machines \cite{basis5}, feature mapping \cite{ham1,var2,ham3,ham4,featuredist}, machine learning problems \cite{angle1,angle2,angle3,angle4}. \\
NISQ compatibility & Partial compatibility: basis encoding, angle encoding and Hamiltonian encoding, due to the current limitations of quantum hardware. \\ \hline 
\end{tabular}
\end{table}

\begin{figure} [!h]
\centerline{\includegraphics[angle = 0,width=1\linewidth]{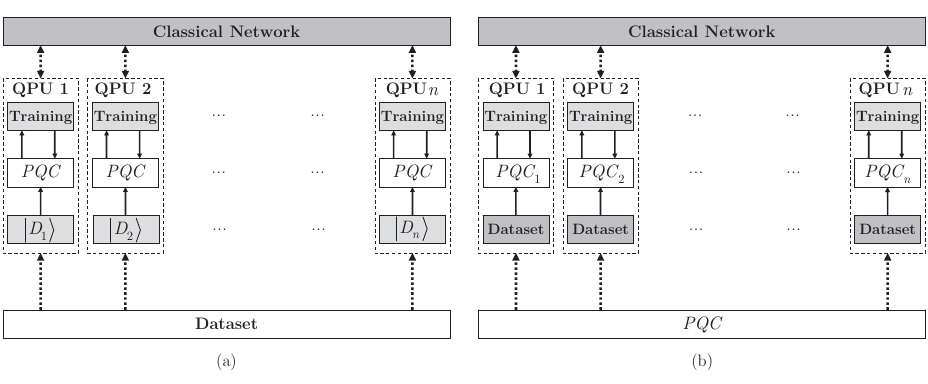}} 
\fcaption{\label{motion} Data splitting and circuit splitting in distributed quantum neural networks. (a) Data splitting. The full dataset (Dataset) is split into $n$ parts, denoted by ${| D_{1} \rangle} ,\ldots ,{| D_{n} \rangle} $, across the $n$ quantum nodes (QPUs), while satisfying $\sum _{i=1}^{n}{| D_{i} \rangle} {\rm \; }={\rm Dataset}$. The model (\textit{PQC}) is loaded into each QPU. The QPUs process and train in parallel, each QPU has access to a classical network (Classical Network). (b) Circuit splitting. The input \textit{PQC} is split into $n$ parts, ${PQC}_{1},\ldots ,{PQC}_{n}$, and loaded into the $n$ QPUs along with the full quantum dataset (Dataset); $\sum\nolimits_{i=1}^{n}{{{c}_{i}}}{PQC}_{i}=PQC$, where ${PQC}_{i}$ is the local PQC of the $i$-th QPU, while ${c}_{i}$ is a real weighting coefficient. The QPUs process and train in parallel; each QPU has access only to a classical network in general.}
\label{dqnn}
\end{figure}

The data splitting and circuit splitting methods in a distributed quantum neural network setting are compared in \fref{dqnn}.

While data splitting is a general strategy for data parallelization in quantum neural networks, other methods have also been investigated. \textit{Data reduction} (or coreset technique) is a different method that can also be used in quantum neural networks to optimize the amount of data shared between the quantum nodes. In data reduction approaches, an approximation of the dataset is distributed between the nodes to improve the parallel processing and to decrease the size of the dataset. In parameter distribution, classical side information is distributed between the nodes in a parallelized way for optimization purposes. \tref{tab4} summarizes the main attributes of data reduction and data distribution in quantum neural networks.

\begin{table}[!h]
\footnotesize
\centering
\tcaption{Data reduction and data distribution approaches in distributed quantum neural networks.}
\label{tab4}
\begin{tabular}{p{1.6in}p{3.1in}} \hline
\centerline{Approach} & \centerline{Description and related works} \\ \hline 
\textbf{Data reduction (coreset)} & Approximation of an original dataset. Application in variational algorithms \cite{reduct1}, and in hybrid quantum-classical architectures\cite{reduct1,reduct2}. \\ 
\textbf{Data distribution} & Parameters are distributed as classical information between the quantum nodes in parallel for optimization \cite{distr1,distr2}. \\
Quantum and classical communication & Quantum communication is available, classical communication is available. \\
NISQ compatibility & Full compatibility. \\ \hline 
\end{tabular}
\end{table}

\begin{figure} [!h]
\centerline{\includegraphics[angle = 0,width=0.5\linewidth]{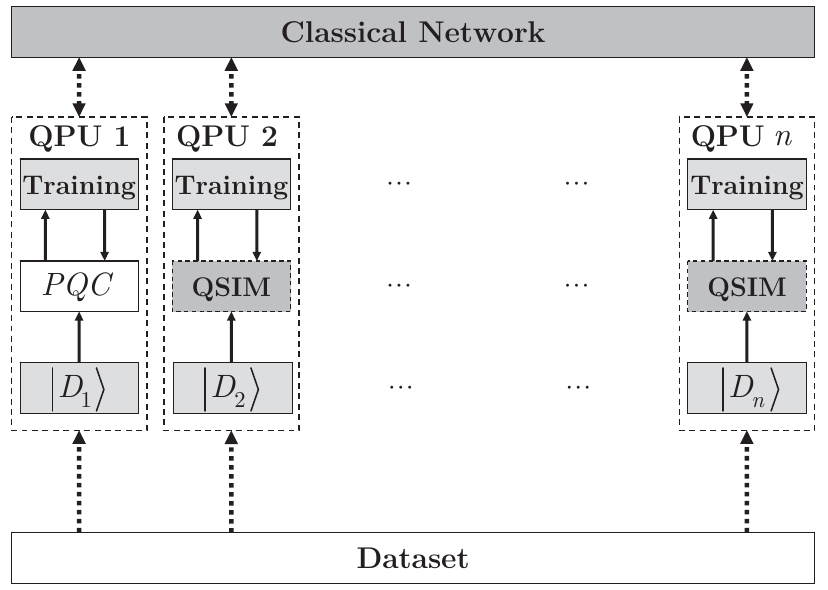}} 
\vspace*{13pt}
\fcaption{\label{motion} A QUDIO model for VQAs. The nodes are trained in parallel using a PQC or a quantum simulator (QSIM). A classical central server communicates with all local quantum nodes to synchronize the trainable parameters of the nodes.}
\label{qudio}
\end{figure}

In \cite{Du}, a modified distributed learning scheme called quantum distributed optimization (QUDIO) has been proposed for variational quantum algorithms. The model consists of a distributed optimization scheme whose back ends support both quantum devices and various quantum simulators to further improve the capabilities of the VQAs. The proposed method collaborates with multiple quantum machines or simulators to complete learning tasks. The model is depicted in \fref{qudio}.

\subsection{Blind Quantum Computation}
In blind quantum computations \cite{blind} multiple users (untrusted devices) perform computations and analysis on the joint data without sharing or revealing their own data with each other. The calculations are hidden from the users and the encrypted input algorithm is received by a server from a trusted client. The data of the users are fed into local quantum devices and a distributed quantum computation is established between them via a particular distributed quantum algorithm \cite{vanmbook, dcomp2}. Classical communication is generally allowed between the quantum devices. A particular user has no knowledge about the other users’ data, nor about the actual distributed quantum algorithm; however, some part of a hidden algorithm and the joint state can be learned by the parties, and the server can also have partial information about some parameters of the calculations. Due to these security issues, another approach for the realization of quantum blind computation is the use of a trusted secure quantum cloud \cite{cloudrev1,cloudadd1b,cloudadd2b}. If the users are connected via a secure quantum cloud, their local quantum devices cannot communicate with each other directly, only via the secure quantum cloud. On the experimental realization of blind quantum computing, see \cite{blinddem1,blinddem2,blinddem3}.

\subsection{Quantum Cloud}
Quantum cloud services allow the users to run quantum algorithms on quantum computers. These cloud services can be accessed via classical computers allowing the users to have access to quantum devices and resources. Several different cloud-based platforms have been developed with different resource capabilities, quantum computing models, and supported software development kits \cite{cloudrev1,cloudrev2,rigettih,cloudadd1,cloudadd2}. In general, the different quantum cloud platforms provide commercial quantum computing as a main service. For a comprehensive review of quantum cloud computing, see \cite{cloudrev1}.

The main attributes of recent quantum cloud platforms with the related works are summarized in \tref{qcloud}. \textit{Superconducting technology:} IBM Quantum\footnote{IBM Quantum Computing Services. URL: \url{https://quantum-
computing.ibm.com/}}, Google Quantum\footnote{Google Quantum AI. URL: \url
{https://quantumai.google/}}, Rigetti\footnote{~Rigetti Computing. URL: \url{https://qcs.rigetti.com}}, OCO: Oxford Quantum Circuits\footnote{~Oxford Quantum Circuits. URL: \url{https://oqc.tech/}}, ~QuTech\footnote{~Quantum Inspire, De Voorhoede. URL:\url{https://www.quantum-inspire.com/}}. \textit{Trapped ion technology:} IonQ\footnote{~IonQ. IonQ Cloud Service. URL: \url{https://ionq.com/}}, ~Quantinuum\footnote{~Quantinuum Cloud Service, Quantinuum. URL: \url{https://www.quantinuum.com/}}. \textit{Neutral atom technology:} Pasqal\footnote{{~Pasqal: Programmable Atomic Arrays. URL: \url{https://www.pasqal.com/}}}, ~QuEra\footnote{{~QuEra, Quantum Computing with Neutral Atoms (2024). URL: \url{https://www.quera.com/}}
}. ~\textit{Quantum annealing technology:} D-Wave\footnote{~D-Wave. URL: \url{https://www.dwavesys.com}}. \textit{~Solid-state spins technology:} QuTech.

\begin{table}[h!]
\centering
\footnotesize
\tcaption{Recent quantum cloud platforms.} 
\label{qcloud}
\begin{tabular}{p{1in}p{0.7in}p{1.6in}p{1.7in}} \hline 
\centerline{Platform} & \centerline{Computing Model} & \centerline{Quantum hardware vendor} & \centerline{Description and related works} \\ \hline
\textbf{IBM Cloud} & Gate-model, quantum simulation & IBM Quantum (1121 qubits) \cite{ibmq} & Remote access to the IBM quantum computing hardware, serverless model \cite{ibmcloud}, supported quantum software: Qiskit. \\ 
\textbf{Google Cloud} & Gate-model, quantum simulation & Google (54 qubits) \cite{googlecloud}, IonQ (36 qubits) \cite{ionq} & Remote access to Google's quantum computing hardware \cite{googlecloud}, supported quantum software: Cirq. \\ 
\textbf{Microsoft Azure Quantum} & Gate-model, quantum simulation & Quantinuum (32 qubits) \cite{nuum}, Rigetti (84 qubits) \cite{rigettih}, IonQ (36 qubits) \cite{ionq}, Pasqal (100 qubits) \cite{pasqal} & Remote access to a diverse portfolio of current quantum hardware \cite{mazure}, supported quantum software: Q\#, Qiskit, Cirq. \\ 
\textbf{Amazon Braket} & Gate-model, quantum simulation & Rigetti (84 qubits) \cite{rigettih}, OCQ (32 qubits), IonQ (36 qubits) \cite{ionq}, QuEra (256 qubits) \cite{quera} & Access to different types of quantum computers, simulators and quantum-classical algorithms, serverless model \cite{amazon,abraket2}, supported quantum software: Braket, Qiskit, Pennylane. \\ 
\textbf{PlanQK} & Gate-model, quantum annealing, quantum simulation & IBM Quantum \cite{ibmq}, Amazon Braket \cite{amazon,abraket2}, Azure Quantum \cite{mazure} & Remote running of quantum tasks and algorithms, provides access to major quantum backends and simulators, serverless model \cite{planqk1,planqk2}, supported quantum software: Qiskit, Pennylane. \\ 
\textbf{QuantumPath} & Gate-model, quantum annealing, quantum simulation & IBM Quantum \cite{ibmq}, Amazon Braket \cite{amazon,abraket2}, D-Wave (5000 annealing qubits), QuTech (5 qubits) \cite{qutech} & Industry-ready hybrid quantum-classical solutions \cite{qpath1,qpath2}, supported software: Qiskit, Ocean, Braket, Q\#. \\ 
\textbf{Strangeworks} & Gate-model, quantum annealing, quantum simulation & IBM Quantum \cite{ibmq}, Amazon Braket \cite{amazon,abraket2}, Azure Quantum \cite{mazure} & Hybrid quantum-classical solutions, serverless model \cite{sworks}, supported quantum software: Qiskit, Braket, Forest. \\ 
\textbf{QFaaS} & Gate-model, quantum annealing, quantum simulation & IBM Quantum \cite{ibmq}, Strangeworks \cite{sworks} & A function-as-a-service framework for quantum computing, open-source, serverless model \cite{qfa1,qfa2}, supported quantum software: Qiskit, Cirq, Q\#. \\ 
\textbf{1Qloud} & Quantum simulation & 1Qbit \cite{1q} & Hybrid quantum-classical solutions \cite{1q}, supported quantum software: 1Qbit. \\ 
\textbf{QEMIST} & Quantum simulation & 1Qbit \cite{1q} & Hybrid quantum-classical solutions \cite{qem}, supported quantum software: OpenQEMIST. \\ \hline    
\end{tabular}
\end{table}

\subsection{Remote Sensing and Interferometry}
Remote quantum sensing uses the fundamentals of quantum mechanics for highly sensitive measurements in a distributed manner via an entangled network of distributed quantum sensors \cite{distrqsens,sens1}. Applications of remote sensing include atomic quantum clocks, quantum sensors for geology, biomedical applications \cite{sens3} and navigation, nanoscale vortex imaging using a cryogenic quantum magnetometer \cite{sens2}, quantum-enhanced accelerometers and gyroscopes, magnetic field detection, and gravitational wave detection.

Quantum interferometry can be also realized in a distributed way using multiple local quantum interferometers at distant quantum computers \cite{interfero, dcomp2}. The local quantum computers process the results of the interferometers and use a distributed quantum algorithm to extract information to improve the accuracy and to decrease the noise level.

\subsection{Further Works}
On the use of QAOA in the NISQ setting, see \cite{qaadd1,f1,qc5,compress,dqaoa}, on variational quantum eigensolvers, see \cite{varadd1,varadd2, varadd3,parvar}, and on pseudorandom quantum circuit-based solutions for quantum advantage, see \cite{pseudo1, pseudo2, pseudo3}. The complexity of NISQ has been derived in \cite{nisqcomp}, while the structure of shallow quantum circuits has been studied in \cite{shallow1,shallow2}. For details on the methods of loading classical information into quantum computers, see \cite{loading}.

For the fundamental details of parameterized quantum circuits, we suggest \cite{qnnimp9,pqc1,pqc2,pqc3,pqc4,pqc5,nisq, pqc7, pqc8}. On the interoperability of quantum neural networks and quantum computations, and on the attributes of the different training procedures of quantum neural networks, we suggest \cite{pqc9,pqc10,pqc11,distr2,pqc12,pqc13}. A quantum-classical hybrid algorithm for determining eigenstate energies in quantum chemistry is proposed in \cite{eigenst}. On resource-efficient quantum algorithms for quantum chemistry, see \cite{variathien}. For an enhanced feature encoding and classification on distributed quantum hardware, see \cite{featuredist}. Machine learning algorithms for near-term quantum computers can also be found in \cite{miadd1,miadd2,miadd3,miadd4,ham1}. For the different learning architectures of quantum neural networks, see \cite{qnnarch1,qnnarch2,qnnarch3,qnnarch4,qnnarch5,qnnarch6, pqc5}. For further works on the barren plateau phenomenon in quantum neural networks, see \cite{qnnbarr1,qnnbarr2,qnnbarr3, pqc3,qnnbarr4,qnnbarr5}.
The implementation aspects of quantum neural networks on current quantum hardware architectures for different problems are studied in \cite{qnnimp1,qnnimp2,qnnimp3,qnnimp4,qnnimp5,qnnimp6, qnnimp7,qnnimp8,qnnimp9}.

\section{Networked Quantum Software, Programming Languages and SDKs}
\label{sec4}
Networked quantum services are supported by different quantum software tools. Several different quantum programming languages are available for writing and running networked quantum services, with different functionalities for preparing the quantum gates, quantum circuits and quantum algorithms \cite{qprog1,qprog2,qprog3,semidef}. The quantum software development kits (SDKs) are integrated frameworks with several different tools, libraries and simulators to develop quantum circuits \cite{qsdk1,qsdk2}. Quantum software libraries (SLs) are also available with functions and tools for development \cite{libr}. The study of Zhao \cite{qsoft1} reviews quantum software engineering, while the study of Murillo et al. \cite{qsoft2} propose a roadmap and reviews the challenges of quantum software engineering.

In a networked quantum computing environment, the quantum software tools and quantum APIs allow to provide different networked quantum services to the users. These services include quantum computing as a service (QCaaS) solutions (such as quantum cloud computing, see \tref{qcloud}), quantum encryption and quantum cryptographic services (such as the QKD services of ID Quantique, MagicQ, etc.), and quantum AI services (AI services of QC Ware, Xanadu, CQC, etc.). 

The limitations of quantum software tools in a networked quantum computing environment are related to the No-cloning theorem (we cannot copy the program states) and the problem of measurement (a measurement destroys the program states). These issues can be handled via quantum simulators \cite{incoh4,survyang} in the stage of quantum software development. Quantum simulators are available in the different quantum SDKs (see \tref{tab5} and \tref{qprog}), while the various quantum simulators and online AI platforms can be connected to serve as an integrated platform for quantum programming and quantum software development \cite{serv2}. The quantum programs then can be tested on real quantum devices with several repetitions and measurement rounds to get a probability distribution of the results.

Quantum programming languages allow the development of quantum operating systems, quantum firmware, and quantum toolkits \cite{qprogadd1} for distributed quantum computations. Quantum software handle distributed quantum computations via error correction protocols, routing protocols, and provide an infrastructure for the various quantum hardware platforms. The networking protocols provided by quantum programming languages allow to adapt and improve the network performance via the enhancement of transmission rates, and the quality of service (QoS) of distributed quantum computations.

A challenge of integrating quantum and classical computing workflows is the cooperation of quantum and classical interfaces and programs. A possible solution to the problem is the utilization of paradigms of service-oriented computing (SOC) in quantum computing. SOC-based approaches allow quantum and classical developers and programmers to collaboratively create hybrid quantum-classical applications in an efficient and scalable way. The works of Kumara et al. \cite{kumara} and Moguel et al. \cite{develop1} focus on this topics. Another challenge is the handling of decoherence and noise in quantum hardware platforms, since disturbances can result in different probability distributions on different quantum hardware. The quantum simulators integrated into the different quantum programming frameworks and quantum SDKs can also result in different classical outputs which makes it difficult to integrate the results.

Relevant attributes of recent quantum software are included in \tref{tab5}.

The attributes of recent quantum programming languages, quantum SDKs and quantum SLs for networked quantum services are summarized in \tref{qprog}.

\begin{table}[!h]
\footnotesize
\centering
\tcaption{Quantum software for networked quantum services.}
\label{tab5}
\begin{tabular}{p{1.1in}p{3.5in}} \hline
\centerline{Quantum Software} & \centerline{Description and related works} \\ \hline 
CutQC & Software package for circuit splitting \cite{incoh6}. \\ 
Interlin-q & Software package for the development of distributed quantum algorithms \cite{interl}. \\ 
Pennylane & Quantum software for simulations and experiments on current NISQ quantum devices, by Xanadu \cite{penny}. \\ 
Qiskit & Quantum software for simulations and experiments on current NISQ quantum devices, by IBM \cite{qis1,libr}. \\ 
QuantumCircuitOpt & Software for the implementation of mathematical optimization and algorithms for decomposing arbitrary unitary gates into a sequence of hardware-native gates\cite{qis1b}. \\
QuPanda & Software for creating and executing complex quantum circuits and algorithms, by Origin Quantum \cite{qupanda}. \\
QuNetSim & Software for real time quantum networks simulation \cite{qunet}. \\ 
Qurzon & A compiler that uses CutQC and other tools \cite{qur}. \\ 
ScaleQC & A software tool for circuit splitting \cite{scaleqc}. \\
SuperSim & A software tool for circuit splitting \cite{supersim}. \\
TorchQuantum & A software for merging quantum computing with deep learning, by MIT \cite{torch}. \\
Tensorflow Quantum & Distributed quantum machine learning \cite{tensor1}, distributed training of quantum neural networks \cite{tensor2,qnnbarr2}, by Google and NASA Ames. \\ \hline 
\end{tabular}
\end{table}

\begin{table}[!h]
\footnotesize
\centering
\tcaption{Quantum programming languages, quantum SDKs and quantum SLs for networked quantum services.}
\label{qprog}
\begin{tabular}{p{1.2in}p{3.5in}} \hline
\centerline{Approach} & \centerline{Description and related works} \\ \hline
\multirow{4}{*}{Programming language} 
& \multicolumn{1}{p{3.5in}}{\textbf{Qiskit:} an integrated programming language and quantum SDK for quantum simulations and quantum algorithms, by IBM \cite{qis1,libr,qisa1,qisa2,qisa3}.} \\
& \multicolumn{1}{p{3.5in}}{\textbf{PyQuil:} a Python library for quantum programming using Quil, the quantum instruction language developed by Rigetti Computing \cite{pyquil}.} \\
& \multicolumn{1}{p{3.5in}}{\textbf{Q\#:} programming language for quantum computing, by Microsoft. Integrated support for different program languages and quantum development kit \cite{qk}.} \\
& \multicolumn{1}{p{3.5in}}{\textbf{Quipper:} an embedded, scalable functional programming language for quantum computing \cite{quipper1,quipper}, by Microsoft and the University of Oxford.} \\ \hline
\multirow{5}{*}{Quantum SDK}
& \multicolumn{1}{p{3.5in}}{\textbf{Ocean:} quantum SDK for quantum annealing algorithms, by D-Wave \cite{dwave1, dwave2}.} \\
& \multicolumn{1}{p{3.5in}}{\textbf{Forest:} quantum SDK for quantum computing using the quantum services of Rigetti \cite{rigetti1,rigetti2}.} \\
& \multicolumn{1}{p{3.5in}}{\textbf{Microsoft QSDK:} quantum SDK for quantum computing using Microsoft quantum services \cite{msdk1, msdk2}.} \\
& \multicolumn{1}{p{3.5in}}{\textbf{Strawberry Fields:} a cross-platform Python library for simulating and executing programs on the quantum photonic hardware of Xanadu \cite{xanadu}.} \\
& \multicolumn{1}{p{3.5in}}{\textbf{ProjectQ:} a Python-based open source framework for quantum computing \cite{projq}.} \\ \hline
\multirow{4}{*}{Quantum SL} 
& \multicolumn{1}{p{3.5in}}{\textbf{Cirq:} a quantum software library by Google for the development of quantum circuits and noise modeling. Cirq provides abstractions for NISQ quantum computers \cite{cirq}.} \\
& \multicolumn{1}{p{3.5in}}{\textbf{OpenFermion:} a quantum software library and electronic structure package for quantum computers \cite{openf}.} \\
& \multicolumn{1}{p{3.5in}}{\textbf{QuTiP:} an open-source Python quantum software library for the dynamics of open quantum systems \cite{qutip}.} \\ \hline
\end{tabular}
\end{table}

\subsection{Quantum API}
A quantum API is a software interface that allows interaction with quantum computers in applications. These APIs also connect the different programming languages and the quantum hardware. The quantum computing vendors provide various APIs (IBM Quantum Experience API, Amazon Braket API, D-Wave API, etc.), therefore a standardization of access to quantum services is an important problem regarding quantum APIs \cite{cloudrev2,api1,api2,api3}. A solution to the standardization problem and for the optimization of accessing quantum services could be the use of quantum API gateways \cite{api1,apigw2}. Quantum API gateways integrate the multiple quantum vendors and offer the best quantum hardware available for a particular task based on the input parameters received from the users’ applications. 

A solution to the problem of service accessing the different platforms is an extension of the quantum API gateway, application of code generators, and a workflow to automate the development and deployment of the services \cite{develop1, develop2}. It allows to deploy quantum code without specific knowledge of the quantum platforms, which can simplify the development of quantum applications and workflow integration. 

An aim of the utilization of programming tools and middleware units is to allow access to quantum services using a standardized format for all quantum providers. A standard format for requesting services and handling service responses also allows developers to manage the communication workflow in an abstract-level, independent from the actual physical quantum hardware and the constraint parameters of the different quantum providers. An approach to standardize the process of defining quantum services using the OpenAPI specification has been proposed in \cite{cloudrev2}. The method can generate source code of quantum services from an API specification and a quantum circuit. For a detailed discussion on the subject of standardization, see the works of Moguel et al. \cite{cloudrev2,survmog}.

The different standardization approaches of service access, service development and programming are summarized in \tref{qapi}. 

\begin{table}[!h]
\footnotesize
\centering
\tcaption{Standardization approaches for networked quantum services.}
\label{qapi}
\begin{tabular}{p{2in}p{2.6in}} \hline 
\centerline{Approach} & \centerline{Description and related works} \\ \hline 
Quantum API Gateway & A machine learning-based middleware for the integration of different quantum vendors for quantum service access \cite{cloudrev2,apigw0,api1,api2,api3,apigw2}. \\
Open API Quantum & A vendor-independent description format for quantum services \cite{opena1,opena2}. \\ 
Qiskit Runtime & API for hybrid quantum-classical computing \cite{cloudrev2}. \\ 
Universal Quantum Intermediate & API for hybrid quantum-classical computing \cite{intermed}. \\ 
Compilation protocol & Simultaneous execution of quantum circuits on NISQ systems \cite{serv1}. \\
Quantum simulators & Connection of different quantum simulators and online AI platforms \cite{serv2}. \\
Cloud computing with load balancing & Quantum-based security approach for cloud computing with load balancing \cite{serv3}. \\
Communication with quantum providers & Quantum secure communication between service provider and subscriber identity module \cite{serv4}. \\
Quantum service prioritization & Integration of different quantum hardware and quantum compilers \cite{prior}. \\
Quality of quantum services & Improving the quality of quantum services via additional layer \cite{qos1,qos2}. \\
Programming of quantum services & Python-based programming in the quantum cloud for quantum services \cite{rigs}. \\
Development and deployment & Tools for the development of quantum services \cite{develop1}, and hybrid quantum-classical services \cite{develop2}. \\
Provenance system & A provenance method for the selection of a quantum computer \cite{prov}. \\
Hybrid quantum applications & Orchestration of quantum and classical services in hybrid systems \cite{orch}. \\ \hline 
\end{tabular}
\end{table}

\section{Implementation Basis}
\label{sec5}
In this section, we review the concepts that are essential for the operation of networked quantum devices, as well as the main types of devices and the requirements.

\subsection{Quantum Channels}
Any two communicating nodes in the system are connected by a quantum channel, a communication medium that is influenced by nature in the form of noise \cite{IGY1}. The theoretical modeling of quantum channels has a long history. Many articles have related the theoretical maximum of the transmission capacity of the channels to their capacity; for understandable reasons, the capacity for classical information has primarily been considered, since in the initial point-to-point systems, the end users were human beings who could only interpret this type of information. However, even in this case, a very diverse range of questions arise; for example, how much classic information can be safely transferred over the channel, and by how much is the capacity increased if we share entangled pairs between the nodes before communication (i.e., when investigating entanglement-assisted capacity)?

The situation becomes completely different when networked quantum devices communicate with each other. This case involves the transmission of quantum information, which is much more sensitive to noise effects, meaning that error correction codes that enable noise control play an even more prominent role. Discussion of this topic is further complicated by the fact that the definitions of the various quantum capacities are not comprehensive, and in some cases there is even a lack of consensus on the quantity describing the information to be maximized.

The capacity of the quantum channel is a guiding quantity, as it maximizes the amount of information that can be transferred. In a practical implementation, error correction codes/algorithms compete for the best approach to capacity \cite{Cai}.

In a similar way to the planning and deployment of classical communication networks, it is also necessary in the quantum case to refine the theoretical channel models based on measurements, to bring them closer to reality.  

According to current trends in quantum communication, there are two parallel directions for future development. Optical fiber-based systems will use existing classical optical communication links that are widely installed worldwide; however, these have a different effect on classical photon packets and the information encoded in the polarization of photons. In a real-world implementation, additional challenges arise if classical and quantum communication travels on the same optical fiber, as artificial interference also inevitably appears in addition to the noise caused by nature.

Quantum free space communication primarily involves the exchange of information with GEO/LEO (geostationary equatorial orbit/low Earth orbit) satellites. As the Earth's atmosphere is relatively thin, its effect on photons is significantly weaker than that of optical fibers, meaning that longer distances can be bridged. This type of communication is also asymmetric, meaning that it is primarily suitable for communication in the satellite-to-Earth direction.

Finally, the role of quantum channel simulators should be emphasized. These are sophisticated tools that are designed to model and analyze the behavior of quantum channels while accounting for various physical effects such as noise, decoherence, and loss. Quantum channel simulators play a crucial role in understanding how quantum states evolve as they are transmitted through the channels, and can help in the development of quantum communication protocols, error correction techniques, and other quantum technologies.

\subsection{Quantum Devices, Quantum Repeaters and Quantum Computers}

When bridging large distances in classical networks, devices may be needed between the endpoints to mitigate losses and distortions in the communication channel. This task is performed by devices called repeaters, which amplify incoming weak signals and send them on. Amplification involves a kind of duplicating (copying) of the incoming weak signal. However, in quantum networks, the No-cloning theorem makes it impossible to copy any incoming arbitrary quantum state without error.

Quantum repeaters are essential components of quantum communication networks, and are designed to extend the range of quantum communication by overcoming the limitations of direct transmission of quantum states \cite{Benchasattabuse}. They are used to facilitate long-distance quantum communication by addressing issues such as loss and decoherence, which occur in optical fibers or free-space communication. 

Quantum long-distance communication uses a combination of two protocols in order to circumvent the No-cloning theorem. Firstly, using entanglement swapping, entangled pairs are created between endpoints located at great distances from each other. Following this, they undergo teleportation, in which quantum states can be transmitted between the parties by communication over a classical channel. The quantum repeaters play an active role in the first phase, as they create entangled pairs between themselves and then perform an operation (Bell measurement) in a synchronized manner, which creates an entangled pair between the endpoints. From the point of view of implementation, the repeater function relies on an extremely precise classical synchronization mechanism between the endpoints. In addition, the resulting entangled quantum states must be preserved for a sufficient time so that they are available for teleportation; in other words, the decoherence time of the quantum memory function must also be sufficiently long.  

Although quantum processors, or computers that have been proven to obey quantum phenomena, have been appearing regularly since 2018 and their physical numbers of qubits are increasing exponentially ($2^n$) every year, meaning that the state space of manageable problems is growing super-exponentially ($2^{2^n}$), there is still competition between different technologies. 

Quantum computers can be categorized into several types based on their underlying technology and operational principles, as follows:
\begin{itemlist}
\item \textit{Superconducting Qubit Computers:} These quantum computers use superconducting circuits to create qubits. Superconducting qubits are among the most widely researched and developed types, with companies like IBM and Google leading the way. They operate at extremely low temperatures to maintain quantum coherence.
\item \textit{Trapped Ion Quantum Computers:} This type uses ions trapped in electromagnetic fields as qubits. Laser beams manipulate the states of the ions. Trapped ion systems, like those developed by IonQ and Honeywell, have demonstrated high fidelity in quantum operations and scalability.
\item \textit{Topological Quantum Computers:} Topological qubits are based on anyons, which are exotic particles that exist in two-dimensional spaces. Theoretical models suggest that topological qubits could be more robust against errors due to their topological nature. Research is ongoing in this area, with Microsoft being a prominent player.
\item \textit{Photonic Quantum Computers:} These systems use photons (light particles) as qubits. Photonic quantum computers can operate at room temperature and utilize optical components like beam splitters and waveguides. They are being explored for their potential in quantum communication and networking.
\item \textit{Quantum Dot Computers:} Quantum dots are semiconductor particles that can confine electrons and create qubits. These systems leverage the properties of quantum dots to perform quantum computations, and researchers are exploring their potential for scalability and integration with existing semiconductor technology.
\item \textit{Neutral Atom Quantum Computers:} Neutral atoms are trapped and manipulated using optical tweezers or magnetic fields. This approach allows for scalable qubit arrays and is being researched for its potential in quantum simulation and computing.
\item \textit{Adiabatic Quantum Computers:} These systems use adiabatic processes to evolve a quantum state from an easy-to-prepare initial state to a hard-to-solve problem state. D-Wave Systems is known for developing adiabatic quantum computers primarily focused on optimization problems.
\end{itemlist}

The wide variety of quantum computers that will be available in the future may limit their networked operation, since only machines in which there is an interface between the qubits performing the calculations and the photons carrying the information will be able to be connected to the network. Another requirement is that quantum computers must be able to preserve quantum stored information for at least as long as the period over which communication between the machines takes place. That is, the coherence time of quantum memories also restricts the range of machines that are capable of cooperation. Here, we draw attention to the possibility that the information storage qubits of quantum memories may be based on a different physical medium. In other words, a convention may be necessary between the qubits that perform calculations, store information, and deliver information; these processes suffer from noise, making the role of the error correction codes even more valuable.

Based on their important roles in quantum communication, we can identify the most promising physical implementations of quantum memory as follows \cite{Navak}:
\begin{itemlist}
\item \textit{Atomic Ensembles:} These systems use collections of atoms or ions to store quantum information. The quantum states of the atoms can be manipulated using light fields, which allows for the storage and retrieval of quantum states. Techniques such as electromagnetically induced transparency (EIT) are often employed in these systems.
\item \textit{Solid-State Systems:} Quantum memories can also be realized using solid-state materials, such as quantum dots or nitrogen-vacancy (NV) centers in diamond. These systems can provide robust storage capabilities, and are often integrated with photonic devices for efficient state transfer.
\item \textit{Photonic Quantum Memories:} These memories utilize photons as carriers of quantum information. Photonic quantum memories can be implemented using nonlinear optical processes, which allows for the temporary storage of photonic quantum states.
\item \textit{Cryogenic Quantum Memories:} Some quantum memory systems operate at cryogenic temperatures to reduce thermal noise and enhance coherence times. These systems often utilize superconducting qubits or other low-temperature phenomena.
\end{itemlist}

In summary, it can be concluded that all quantum devices intended for use in network operation must meet three requirements: \textit{precise synchronization}, \textit{efficient error-correction coding} and \textit{memory with adequate coherence time}. These functions do not necessarily require different physical implementations, and combining them can increase efficiency.

\subsection{Entanglement-Assisted Networked Computing} 
In networked quantum services, a particular quantum state can be affected by noise from various sources such as quantum gate errors, channel losses, environment noise. The $F$ fidelity of the state, $0\le F\le 1$, can be used to quantify how much a particular state has been affected by these noise sources \cite{vanmbook, archprinc}. The generation of Bell pairs, or the process of entanglement swapping are both noisy operations, and as a corollary it degrades the fidelity of a quantum state. Higher-fidelity Bell states can be generated from lower-fidelity pairs via the process of entanglement purification (or entanglement distillation). Quantum error correction (QEC) can also be utilized in entanglement distillation with lower communication costs \cite{dur2007}, but with higher demands on the fidelity of the input quantum state. The resulting output fidelity, $F'$, of the entanglement purification at an input state fidelity $F_0$ is $F'=F_0^2/p_S$, with relation $F'>F_0$ if $F_0>0.5$, where $p_S$ is the success probability of entanglement purification, $p_S=F_0^2+(1-F_0)^2$. The utilization of Bell pairs in networked quantum services depends on a $F_{\rm{thr}}$ fidelity threshold, below which the entangled states cannot be used in practice. Different applications can have different requirements on $F_{\rm{thr}}$. In networked quantum services, a Quality of Service (QoS) parameter can be the end-to-end Bell pairs per second (BPPS) and the required fidelity of the Bell pairs \cite{survweh, vanmbook, wang}.

Scalability of networked services requires the extension of quantum systems over long-distances while preserving quantum coherence. The current hardware limitations of quantum devices such as gate fidelities and error rates makes the problem of scaling even more difficult \cite{scal,scal1,scal2,scal3}. The handling of external disturbances, unwanted interactions and decoherence in quantum systems are the main challenges in scaling networked quantum services. To achieve scalability over long distances in networked quantum services, the quantum repeater architecture of the quantum internet can be utilized with advanced quantum networking protocols \cite{cacm}. An integration of physical layer developments (entanglement swapping, entanglement purification, multipartite entanglement, quantum error correction, quantum memories, controlling) with higher layer advancements (entanglement routing, delay minimization techniques, dynamic path selection, network intelligence, error management, connectivity management, etc.) provides an effective solution for the scalability of practical networked quantum services.

\subsection{Quantum Error Correction, Fault-Tolerance, Quantum Memories}
The aim of quantum error correction is to preserve quantum information decoherence and errors. By using quantum error correction codes (such as surface codes, error syndromes, code concatenation) the accuracy and reliability of quantum information processing can be enhanced significantly. The tools of quantum error correction allow us to realize fault-tolerant quantum computations in experiment. According to the threshold theorem \cite{errorprop4} (or quantum fault-tolerance theorem), a quantum computer with a physical error rate below a certain threshold can suppress the logical error rate to arbitrarily low levels via the application of quantum error correction. This means that quantum computers can be made fault-tolerant. 

Due to the efficient error correction techniques, the percentage of error of quantum gates is decreasing over the years \cite{errors,attya,qcadd12}. The current $\varepsilon $ error rate (\%) of quantum gates is in the range of $\varepsilon \approx 0.01$, which represents approximately two orders of magnitude improvement in the last few years. \tref{gaterror} summarizes the $\varepsilon $ gate error rates of quantum processors of different quantum vendors.

\begin{table}[h!]
\footnotesize
\centering
\tcaption{Gate error rates of quantum processors.}
\label{gaterror}
\begin{tabular}{p{1.5in}p{1in}p{0.5in}} \hline 
\centerline{Quantum processor} & \centerline{Error rate (\%)} & \centerline{Release date} \\ \hline 
IBM Quantum Eagle & \centering {$\sim $0.9} & 2019\\
Google Sycamore & \centering {$\sim $0.6} & 2020\\
Rigetti Aspen-9 & \centering {$\sim $0.5} & 2021\\
IonQ Harmony & \centering {$\sim $0.3} & 2022\\
IBM Quantum Hummingbird & \centering {$\sim $0.2} & 2023\\ 
Google Quantum Bristlecone & \centering $\sim $0.15 & 2023\\
Rigetti Aspen-12 & \centering $\sim $0.1 & 2024\\
IonQ Symphony & \centering $\sim $0.01 & 2024\\\hline
\end{tabular}
\end{table}

An advantage of networked quantum services is the \textit{enhanced fault-tolerance} and \textit{improved scalability} due to the distributed architecture \cite{survcal}. For a detailed discussion on quantum error correction, fault tolerance, and quantum memory limitations, we suggest references \cite{error1,error2,error3,error4}.

In networked quantum services, quantum error correction can be implemented with a high cost due to the conditions on physical resources (high number of physical qubits with high fidelity). According to a current roadmap \cite{archprinc} of quantum networks, quantum error correction will be used only in later generations of quantum networks, while current networked quantum services can use entanglement swapping and quantum teleportation. Using Bell states with small-scale error correction such as entanglement distillation represent an implementable alternative of high-cost quantum error correction schemes in practical scenarios. 

Quantum networks can be categorized into Types I-III, based on their loss and error tolerance schemes \cite{murald, archprinc}, as it is summarized in \tref{errorm}. 

\begin{table}[h!]
\footnotesize
\centering
\tcaption{Error correction and management in quantum networks.}
\label{errorm}
\begin{tabular}{p{0.5in}p{1.3in}p{1.3in}p{0.5in}p{0.7in}} \hline 
\centerline{Network type} & \centerline{Loss tolerance scheme} & \centerline{Error tolerance scheme} & \centerline{Delay} & \centerline{Cost}\\\hline 
Type I. & Heralded entanglement generation with bidirectional classical side-information & Entanglement distillation with bidirectional classical side-information & High & Polynomial scaling with total distance\\
Type II. & Heralded entanglement generation with bidirectional classical side-information & Entanglement distillation with unidirectional classical side-information, or QEC with no classical side-information & Moderate & Polylogarithmic scaling with total distance\\
Type III. & QEC with no classical side-information & QEC with no classical side-information & Low & Polylogarithmic scaling with total distance
\\\hline
\end{tabular}
\end{table}

Type I quantum networks use heralded entanglement generation, the entangled states are probabilistically generated and the distribution range is extended via entanglement swapping. The term "heralded" refers to that the entanglement generation process runs until the two non-neighboring stations do not receive an ACK (acknowledge) message from the intermediary station, which confirms that the entanglement swapping operation was successful. High-fidelity entanglement is produced via entanglement purification. These networks use two-way classical communications which reduces significantly the achievable communication rates.

From Type II quantum networks, the entanglement purification procedure can be replaced by QEC. It improves the rate of communication since it can reduce the required amount of classical communications. However, QEC also requires higher state fidelities and additional quantum circuits that makes the scheme more complex. 

Type III quantum networks use QEC to handle the loss and operational errors without classical communications. These networks can use direct unidirectional communication, but with an improved complexity due to the use of QEC for loss and error tolerance. Type III networks can directly transmit QEC-encoded qubits to adjacent quantum nodes, and the Bell pairs can be created without heralding or the use of classical side-information \cite{archprinc}, however they require high-fidelity quantum gates and longer coherence times.

\tref{repqec} summarizes the applications of different QEC codes \cite{errors} in Type II-III networks.

\begin{table}[h!]
\footnotesize
\centering
\tcaption{Error correction in Type II-III quantum networks.}
\label{repqec}
\begin{tabular}{p{0.5in}p{2in} p{2in}} \hline 
\centerline{Network type} & \centerline{QEC code} & \centerline{Application}\\\hline 
Type II. & Repetition code, Shor code, Calderbank-Shor-Steane (CSS) code & Correction of operational errors. \\
Type III. & Surface code, Gottesman-Kitaev-Preskill (GKP) code & Correction of photon
loss and operational errors.\\\hline
\end{tabular}
\end{table}

Regarding quantum memories, a critical problem in achieving persistent storage is the isolation a quantum system from the environment. Due to the decoherence, the environment adds an uncontrollable noise into the system, which also degrades the fidelity of the state. The lifetime (coherence time) of a quantum memory is implementation-specific, however the highest achievable values currently in a quantum network hardware are in the range of few seconds to few minutes (see \tref{qmem}). The coherence time values of quantum memories are continuously increasing as the different technologies evolve. Networked quantum services also have to be able to manage short coherence times. A possible solution to this problem is the reduction of the latency on critical paths of the network \cite{archprinc}. Details of recent quantum memory implementations with their highest coherence time values are given in \tref{qmem}.

\begin{table}[h!]
\footnotesize
\centering
\tcaption{Quantum memory implementations and their lifetime (coherence time) values.}
\label{qmem}
\begin{tabular}{p{2.5in}p{1.2in}} \hline 
\centerline{Quantum memory} & \centerline{Coherence time} \\ \hline 
Single ion qubit \cite{qmem2} & $\sim $1 hour\\
Single trapped ion qubit \cite{qmem1} & $\sim $10 min\\
Ten-qubit solid-state spin register \cite{qmem5} & $\sim $1 min\\
Single electron spin coupled to a multi-qubit nuclear-spin environment \cite{qmem4} & $\sim $1 sec\\
Superconducting cavity qubit \cite{qmem3} & $\sim $Tens of milliseconds\\\hline
\end{tabular}
\end{table}

\section{Conclusion and Outlook}
\label{sec6}

The advances in quantum computation and communications provide the ground for novel developments such as networked quantum services, which allow users to access quantum resources and efficient solutions in quantum networking scenarios and in hybrid quantum-classical architectures. In a similar way to classical systems, the networked operation of quantum devices will bring qualitative advances, representing the third quantum revolution. This paper has reviewed the topic of the networking of quantum devices, which requires significant theoretical and technological progress. In our article, we have presented the gate-model concept and its potential applications in various fields of the quantum world. We have also pointed out the technological requirements that networked devices must meet. 

From the point of view of future developments, significant challenges remain from both the theoretical and practical/technological points of view. Networked communication is currently at the experimental, test network TRL (technology readiness level) stage. In the future, highly scalable versions of these solutions will be needed, and the overall capabilities of network devices must also be enhanced. The importance of ensuring interoperability between devices from different manufacturers cannot be overemphasized, and to achieve this, it will also be necessary to harmonize the parallel standardization efforts.

Further open problems related to networked quantum services are the improvement of the accessibility of quantum services, the automatic generation and deployment of quantum services, as well as novel service engineering methods for service developers to simplify the development and deployment of quantum services. Another important direction for future research is the API standardization and middleware development to access quantum services of different service providers in a simplified and standardized way, and the development of a standardized architecture stack. The interoperability of classical and networked quantum services is another important subject, along with the development of hybrid high-performance quantum-classical networks. The integration of quantum key distribution and quantum cryptographic primitives into networked quantum services and quantum-classical architectures is another direction for research. Regarding quantum software of networked services, an important task is the development of a standardized programming language for the services. A current challenge for quantum software is the application of distributed algorithms and compilers. Most of the hardware challenges of networked quantum services coincidence with the challenges of the NISQ era devices. These problems are centered around decoherence and noise reduction, limited capabilities and lack of error correction. An open research problem is the efficient implementation of networked quantum services via actual noisy quantum devices and shallow quantum circuits. By using intermediate quantum repeaters, the distant quantum devices can be coupled. However, the quantum hardware and quantum resources are limited by the capabilities of the current quantum devices, which cannot provide high performance generation and distribution of entanglement between the distant nodes. Another challenge is the handling and management of the noise of the quantum links and the reduction of communication overhead between current quantum devices. An important research problem is the integration of different enabling technologies to quantum devices to improve the performance of quantum networking, and the establishment a scalable architecture of networked quantum services via the quantum internet.

This survey provided an overview of the concepts and attributes of networked quantum services. We studied the architectural background, the different applications, the quantum software and quantum programming languages, and the approaches to the standardization of quantum APIs and service accessing. The survey presented a comprehensive review of the state of the art of networked quantum services in a compact form. With this research, we expect to contribute to the further development of the field of networked quantum services.

\nonumsection{Acknowledgements}
\noindent
This research was supported by the Ministry of Culture and Innovation and the National Research, Development and Innovation Office within the Quantum Information National Laboratory of Hungary (Grant No. 2022-2.1.1-NL-2022-00004).

\nonumsection{References}

\end{document}